\documentclass[sigconf]{acmart}
\usepackage{booktabs} 

\setcopyright{rightsretained}

\usepackage[utf8]{inputenc}

\usepackage{amsmath}
\usepackage{amsthm}
\usepackage{amssymb}
\usepackage{enumerate}
\usepackage{algorithm}
\usepackage{algpseudocode} 
\usepackage{xspace}
\usepackage{multirow,multicol}
\usepackage{framed}
\usepackage{array}
\usepackage{chngcntr}

\usepackage[shortlabels]{enumitem}

\usepackage{etoolbox}
\patchcmd{\paragraph}{\@parfont}{\bfseries}{}{} 
\patchcmd{\paragraph}{\parindent}{0pt}{}{}

\usepackage{amsthm}
\usepackage{calc}

\usepackage{graphicx}

\usepackage{xcolor}

\newcolumntype{P}[1]{>{\centering\arraybackslash}p{#1}}
\newcolumntype{M}[1]{>{\centering\arraybackslash}m{#1}}

\copyrightyear{2017} 
\acmYear{2017} 
\setcopyright{licensedusgovmixed}
\acmConference{ACSAC 2017}{December 4--8, 2017}{San Juan, PR, USA}\acmPrice{15.00}\acmDOI{10.1145/3134600.3134609}
\acmISBN{978-1-4503-5345-8/17/12}

\author{Adam J. Aviv}
\affiliation{United States Naval Academy}
\email{aviv@usna.edu}

\author{John T. Davin}
\affiliation{United States Naval Academy}
\email{john.t.davin@gmail.com}

\author{Flynn Wolf}
\affiliation{University of Maryland, Baltimore County}
\email{flynn.wolf@umbc.edu}

\author{Ravi Kuber}
\affiliation{University of Maryland, Baltimore County}
\email{rkuber@umbc.edu}

\title{Towards Baselines for Shoulder Surfing on Mobile Authentication}

\begin{document}

\begin{abstract}
  Given the nature of mobile devices and unlock procedures, unlock authentication is a prime target for credential leaking via
  shoulder surfing, a form of an observation attack. While the research community has investigated solutions to minimize or prevent the threat of shoulder surfing, our understanding of how the attack performs on
  current systems is less well studied. In this paper, we describe a
  large online experiment ($n=1173$) that works towards establishing a
  baseline of shoulder surfing vulnerability for current unlock
  authentication systems. Using controlled video recordings of a
  victim entering in a set of 4- and 6-length PINs and Android unlock
  patterns on different phones from different angles, we asked
  participants to act as attackers, trying to determine the
  authentication input based on the observation. We find that 6-digit
  PINs are the most elusive attacking surface where a single
  observation leads to just 10.8\% successful attacks (26.5\% with multiple observations). As a comparison, 6-length Android
  patterns, with one observation, were found to have an attack rate of 64.2\% (79.9\% with multiple observations). Removing feedback lines for
  patterns improves security to 35.3\% (52.1\% with multiple observations).  This evidence, as well as
  other results related to hand position, phone size, and observation
  angle, suggests the best and worst case scenarios related to shoulder
  surfing vulnerability which can both help inform users to improve
  their security choices, as well as establish baselines for
  researchers.
\end{abstract}


\maketitle

\section{Introduction}
Personal and sensitive data is often stored on or accessed via mobile devices, making
these technologies an attractive target for
attackers~\cite{egelman2014areyouready}. In the physical domain, the
first line of defense against a proximate attacker seeking to gain
access to the device is the unlock authenticator, the method used to
authenticate the device owner to the device, e.g., by entering a
4-digit PIN.

One type of attack faced when authenticating via a mobile device is shoulder surfing, a form of an
observation attack by which an attacker attempts to observe the
authenticator of a victim while the authenticator is being entered on
the device~\cite{wiedenbeck2006design}. One of the most cited dangers
for smartphone unlocking mechanisms are shoulder surfing
attacks~\cite{man2003shoulder}.

While many users utilize biometric authentication as a supplement to
the dominant PIN and graphical (stroke-based) pattern password entry
mechanisms, this does not provide universal protection from shoulder
surfing. Biometrics are a promising advancement in mobile
authentication, but they can be considered a reauthenticator or a
secondary-authentication device as a user {\em is still required to
  have a PIN or pattern} that they enter rather frequently due to environmental impacts (e.g., wet hands).  There are also known to be high false negatives rates associated with
biometrics~\cite{bhagavatula2015biometric}. Further, users with
biometrics often choose weaker PINs as compared to those
without~\cite{cherapau2015impact}, suggesting that the classical
unlock authentication remain an important attack vector going forward.

There is much related work that both proposes and studies shoulder
surfing resistant authentication
mechanisms~\cite{egelman2014areyouready, forget2010eyegaze,
  man2003shoulder, deluca2014nowyouseeme, deluca2009look,
  deluca2012touchme, deluca2010colorpin,
  kumar2007reducing,gao2010new}, but {\em research related to
  understanding the susceptibility to shoulder surfing of currently
  used unlock authentication, namely PINs and Android graphical
  pattern unlock, is limited in nature.} Further, as researchers propose methods and
authentication schemes that offer protections from shoulder surfing
attacks, we do not have clear baselines of comparison for improvement
(or lack thereof) to current schemes.

In this paper, we report the results of a comprehensive study of
shoulder surfing based on video recordings of a victim
authenticating. Our participants, upon viewing the videos, were asked to
recreate the authentication sequences, simulating basic shoulder surfing
attacks. While prior work has considered visual observations of Android
graphical passwords, such as smudge attacks~\cite{aviv2010smudge} and
animated tracing~\cite{vzw2015easy}, prior research only considered a single
dimension. We attempt to account for multiple conditions. 
\begin{itemize}
\item Authentication Type: we compared 4- and 6-digit PINs, and 4- and
  6-length Android graphical patterns, with visible line feedback and without.
\item Observation Angle: we considered 5 different observation angles
  based on videos recorded simultaneously during authentication.
\item Repeated Viewing: we consider situations where the participant
  has a single view of authentication or multiple views.
\item Phone Size: we consider two different touchscreen sizes that are
  common in today's market.
\item Hand Position: we considered two different hand positions to interact with the device,
  single handed thumb input, and two handed index-finger input.
\end{itemize}

We constructed a comprehensive web-based survey and recruited
participants locally from our institution ($n=91$) and online via
Amazon Mechanical Turk ($n=1173$) for a mixed-factorial subject
study. Participants, acting as attackers, were presented with a set of
randomized conditions and asked to view a video of an authentication.  They then attempted to recreate the authentication. 

Analyzing the results, we find that in all settings, Android's
graphical pattern unlock is the most vulnerable, especially when
feedback lines are visible; a single observation successfully
attacked the pattern 64.2\% of the time with 79.9\% for multiple
observations of a 6-length pattern. Shorter patterns were even more
vulnerable. Removing feedback lines during the pattern entry improved
the security, finding 35.3\% successful attacks with a single view and
52.1\% success with multiple views for 6-length patterns (confirming prior
work~\cite{vzw2015easy}). PINs, however, proved much more elusive to
attack than anticipated. A single observation was sufficient to attack
just 10.8\% of the 6-digit PINs, degrading to 26.5\% after two
observations.

These results support what we as a community have believed to be true
anecdotally, and further demonstrates that current authentication
methods provide stronger security against shoulder surfing than one
might expect. Further, these results suggest that baselines of
shoulder surfing success can be applied to this space, to better
support mobile authentication users.  Future work should allow for
improvements over the current worst settings and best settings for
shoulder surfing.




\section{Background and Related Work}

\begin{figure}[t]
\centering
\fbox{\includegraphics[width=0.25\linewidth]{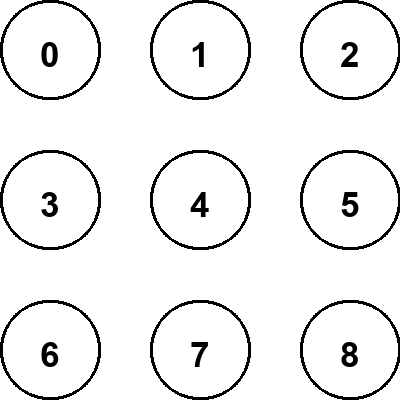}}
\hspace{+.2in}
\fbox{\includegraphics[width=0.25\linewidth]{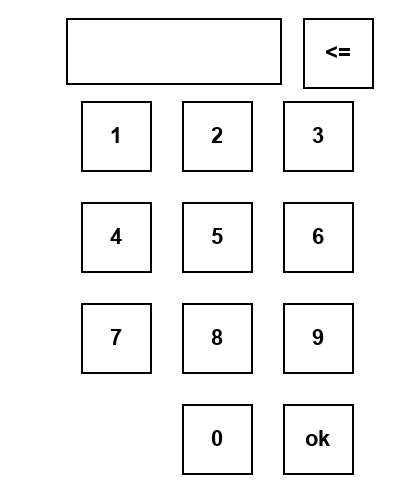}}
\caption{Pattern contact points (left), with label indexing beginning
  at 0, ending at 8, and a PIN layout (right) with digits 0-9, an OK button,
  backspace button and display screen}
\label{fig:samples}
\end{figure}

\paragraph{Mobile Authentication Unlock Choices}
In order to secure access to a mobile device, users are able to use three main mechanisms to unlock the screen.  

\begin{itemize}

\item PIN based authentication, sometimes referred to as a passcode
  on iPhones: where a user is asked to recall a PIN of a least four
  digits. Newer iPhones, however, require a 6-digit passcode
  \cite{iphone6digit}. (A sample PIN layout, as used in our experiments, is shown in Figure~\ref{fig:samples}.)

\item Pattern based authentication: where a user is asked to recall a
  gesture that interconnects a set of 3x3 contact points. On the
  Android OS, four or more points should be selected. The user must
  maintain contact through the authentication, may not reuse contact
  points, nor jump over points previously un-contacted in connecting
  two points. Figure~\ref{fig:samples} shows the grid layout for
  patterns, as well as our labeling scheme of starting with index 0
  through 8. For example, the L-shaped pattern would be 03678. 

  Additionally, pattern based authentication occurs in two
  flavors. The traditional setting is that visual feedback line is
  displayed as the user traverses the contact points (so called,
  with-lines).  The second version requires the user to do the same
  input, but the feedback or tracing lines are not displayed (so called,
  without-lines). In prior work, it has been suggested that the without-lines version of
  the Android pattern is more secure from observation attacks, like
  shoulder surfing~\cite{vzw2015easy}.

\item Password based authentication, sometimes called an alpha-numeric
  passcode: where a user is asked to recall a standard text-based
  password (entered using a soft-keyboard) to unlock the device.

\end{itemize}

The usability and security of PINs
\cite{vonZezschwitz2013wild,FC:BonPreAnd12}, patterns
\cite{uellenbeck, aviv2015isbigger} and passwords
\cite{melicher2016mobile,SP:KKMSVB12} have been well documented by
researchers. Beyond these methods, picture-based
\cite{zhao2013security} and biometric based mechanisms are also used
to unlock mobile devices. The latter is becoming more
commonplace. Fingerprint readers (e.g. TouchID on iPhone v5 or later)
and face identification (through apps such as FaceCrypt or
FastAccessAnywhere) can be used to verify the identity of the
user. Biometrics are often utilized as a secondary authentication
method, and a user with biometric authentication enabled must also
have a PIN set. While biometrics offer promise to promoting quick
authentication, threats related to spoofing and the vulnerabilities
associated residual information left on sensors by victims can pose
challenges to users \cite{tipton2014ios}. In this study, we focus on
two most widely used authentication
mechanisms~\cite{egelman2014areyouready,harbach2016anatomy}, PINs and
graphical Android patterns, with- and without feedback lines.

\paragraph{Shoulder surfing vulnerability}
Numerous types of attacks exist where mobile authentication sequences can be obtained and used by third parties (e.g., simple guessing, smudge attacks, malware attacks~\cite{egelman2014areyouready}). Mobile device users are particularly susceptible to observational attacks, as these devices are used in a range of public and unfamiliar environments where threats may be present. Inputs can be observed and recreated. Furthermore, accessibility features such as magnification of the typed character or displaying the last typed character as cleartext in the password entry field may compromise security~\cite{schaub2012password}. 

Attacks may be performed through direct observation (potentially enhanced through binoculars or low-power telescopes), or through the use of recording devices (e.g. video cameras for later playback) which can be used to covertly obtain or infer credentials \cite{kim2011new,wiedenbeck2006design}. Even if the user attempts to shield the screen from onlookers, security may be compromised through eavesdropping; listening to secure information which can later be used for purposes of recreating entry to a mobile device.  Research reveals that human adversaries, even without recording devices, can be more effective at eavesdropping than expected, in particular by employing cognitive strategies and by training themselves \cite{kwon2014covert}.


\paragraph{Mechanisms to minimize occurrences of shoulder surfing}
Solutions to reduce shoulder surfing include methods of obscuring
entry (e.g., through the use of screen filters, such as Amzer Privacy
Shield, described by \cite{zakaria2011shoulder}), limiting the ability
of third parties to view authentication stimuli input from a specific
angle. Drawmetric solutions also exist where input is made on the back
and/or on the front of the mobile touchscreen device, obscuring the
onlooker's view (e.g. the XSide system
\cite{deluca2014nowyouseeme}). Other drawmetric approaches utilize
behavioral biometrics, which can provide an additional authentication
factor, to verify the user \cite{van2017draw}.

Decoy or randomization scenarios have also been
proposed~\cite{zakaria2011shoulder,von2015swipin}, where, even after
an observation, it challenges observers in recreating the
authentication because he/she cannot differentiate between true and
random input. Touch sensitivity can also be effective. A prescribed
level of pressure during input is difficult for an attacker to
recreate~\cite{malek2006novel}. Similarly, unobservable, tactile
feedback can also be used to thwart a shoulder
surfer~\cite{de2009vibrapass, ali2016developing, krombholz2017may}, where the device
informs the user which of a set of passwords (or nonces) to expect.
 
Kim et al. \cite{kim2011new} suggest that current approaches to
reducing shoulder-surfing typically also reduce the usability of the
system; often requiring users to use security tokens, interact with
systems that do not provide direct feedback or require additional
steps to prevent an observer from easily disambiguating the input to
determine the password~\cite{roth2004pin,
  wiedenbeck2006design}. Bianchi and Oakley
\cite{bianchi2014multiplexed} suggest that authentication becomes a difficult, challenging task as some systems targeting
security against malicious attackers typically place high demands on users.  Wiese and Roth \cite{wiese2015pitfalls} highlight the difficulties in ascertaining the efficacy of shoulder-surfing-resistant technologies due to the lack of comparative studies.  The researchers have highlighted that as set-ups and assumptions made vary by author, it can be difficult to determine the security and usability of
solutions. 

Additionally, most of these studies do not compare directly to the
current state of the art in mobile authentication, namely, how well do
PINs or patterns (or other current methods) perform under attack. We
attempt to fill in that gap here by providing some baselines for what level of security to expect from current authentication choices.

\paragraph{Evaluations of shoulder surfing using video recordings}
According to \cite{wiese2015pitfalls}, in order to determine the
resistance of an interface to shoulder-surfing, the three main methods
used by researchers include: (a) participants are cast into the roles of
adversaries and users, where adversaries observe authentication sessions
of users; (b)  an expert adversary observes the
authentication sessions of all participants; or, (c) participants
are cast into the role of adversaries and observe authentication
sessions of an expert user. While each method has its own advantages
and disadvantages, considerations should be made regarding learning,
motivation and aptitude, to develop a more reliable perception of
risk.

Most related to this work is when researchers present participants
with sets of video recordings depicting actors attempting to
authenticate entry.  Recordings generally aim to simulate an over the
shoulder view. Setting up the videos in this manner ensured that the
attackers would not be affected by inconsistency caused by the
target~\cite{deluca2014nowyouseeme, sherman2014user}.  In prior work,
the choice of number of observations appears to be arbitrary in
nature~\cite{wiese2015pitfalls}. 
Schaub et al. \cite{schaub2013exploring} aimed to
determine how participants fare when attempting to recreate
authentication sequences, comparing those watching video footage vs
live attempts (i.e., physically viewing over a user's shoulder).
Findings revealed that the success rate of video observations are
lower for almost all schemes than the respective live results, with a
few exceptions deviating by only 1–2 observations. Wiese and Roth
\cite{wiese2015pitfalls} recommend preferring live observations to
study human shoulder surfing unless good reasons speak in favor of using
video.

As our study attempted to perform a large-scale, controlled study to systematically compare the two authentication methods, we opted to use video recordings of a single ``expert user'' being attacked by our participants. This allowed us to perform finely tuned randomizations and make comparisons between conditions. As such, as suggested by prior
work~\cite{schaub2013exploring,wiese2015pitfalls}, one can consider
these results as lower-bounds on the security. Live observations from
the same angle would likely increase the vulnerability to shoulder surfing.

\paragraph{Baselines and guidance}
While researchers have extensively explored ways to address
shoulder-surfing attacks, recommendations have been proposed on ways
to design and conduct these types of study.  For example, Wiese and
Roth \cite{wiese2015pitfalls} recommend rather than arbitrarily
selecting a number of observations, that the number of observations
made by adversaries should match their assumptions about the scenario
and the environment where the scheme will be deployed. Observation
strategies should also be taken into account, to gain a more detailed
view of feasible strategies. In terms of set-up, Sahami Shirazi et
al. \cite{sahami2012assessing} propose recording video footage from
four different angles: front, rear, left and right, in order to
compensate the loss of 3D information in 2D videos. While limited
detail was provided about the relative positioning of each camera,
this type of technique would be useful to better simulate shoulder
surfing scenarios.  Schaub et
al. \cite{schaub2013exploring} have highlighted different ways that
users hold and interact with the device. Occlusion by the user’s hand
and fingers may reduce visibility for shoulder surfers and enhance
observation resistance.  

As we will describe in the next section, we attempt to account for
many of these factors and suggestions. Namely, we apply video
recordings from multiple angles, allow for repeated observations and
repeated entries, and we also consider different form factors and hand
positions for our mobile devices.




\section{Methodology}
We designed a mixed-factorial design with both between- and
within-subject factors in order to reduce the duration of the study to an
acceptable length. Between subjects, we randomized participants into
12 groups based on the authentication type (3-treatments), hand
position (2-treatments), and phone type (2-treatments). Within each
group, participants were shown a series of videos for a set of 10
authentications. After each video, each participant attempted to
recreate the authentication observed. As part of a within group
analysis, the observation angles, the number of observations, and the
number of attempts to recreate the authentication were randomized.

Based on this design, we intended to address the following set of
hypotheses:
\begin{itemize}
\item {\bf H1}: The type of unlock authentication, PIN, Pattern
  with-lines, Patterns without-lines, affects the shoulder surfing
  vulnerability.
\item {\bf H2}: Repeated viewings of user input increase the
  likelihood of a shoulder surfing vulnerability.
\item {\bf H3}: Multiple attempts to recreate the input increase the
  likelihood of a shoulder surfing vulnerability.
\item {\bf H4}: The angle of observations affects shoulder surfing
  vulnerability.
\item {\bf H5}: The properties of the unlock authentication, such as
  length and visual features, affect shoulder surfing vulnerability.
\item {\bf H6}: The phone size affects shoulder surfing vulnerability.
\item {\bf H7}: The hand position used to hold and interact with a device affects shoulder surfing vulnerability.

\end{itemize}
In the remainder of this section, we outline the settings of our
experiment and the design choices made. We first discuss the settings of
our video recordings that dictate the participant groups, following which we
discuss the password/PINs used in the experiments, how they were
selected, and the properties they exhibit. Finally, we discuss the
survey mechanisms, training, and other procedures.

\subsection{Video Recording Settings}

\begin{figure}
\centering
\small
\begin{tabular}{c c}
\multicolumn{2}{c}{{\bf Nexus 5} ({\em red phone})}\\
\multicolumn{2}{c}{5.43''x2.72'' form factor, 4.95'' display, 1080x1920} \\
\includegraphics[width=0.25\linewidth]{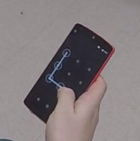} &
\includegraphics[width=0.25\linewidth]{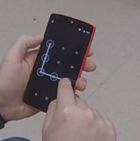} \\
\textit{Thumb} & \textit{Index} \\
\\
\hline
\\
\multicolumn{2}{c}{{\bf OnePlus One} ({\em black phone})}\\
\multicolumn{2}{c}{ 6.02''x2.99'' form factor, 5.5'' display, 1080x1920} \\
\includegraphics[width=0.25\linewidth]{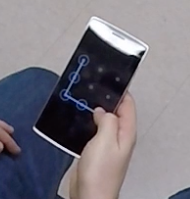} &
\includegraphics[width=0.25\linewidth]{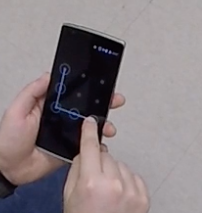} \\
\textit{Thumb} & \textit{Index} \\
\end{tabular}
\caption{Phone Types and Hand Positions: {\em top} is the Nexus 5x phone and {\em bottom} is the OnePlus One phone. The Nexus 5x is roughly the size of a iPhone 6s and the OnePlus one is roughly the size of a iPhone 6s+. On the {\em left} is single handed entry, using the thumb only, and on the {\em right} is two handed entry using the index finger. }
\label{fig:phones} 
\vspace{-.2in}
\end{figure}

\paragraph{Phone settings}
We used two phones in our experiments: Nexus 5 and the OnePlus
One. The Nexus 5 is a mid-range size phone, with a 5'' display. The
OnePlusOne has a larger form factor of 6'' (compared to 5.4'' of the
Nexus 5) with a screen size of 5.5''. Both phones have the same
resolution of 1080x1200 pixels. These two phones are similar to a wide
variety of displays and form factors available on the market today,
for both Android and iPhone. In charts and tables, we refer to the
phones by their coloring, {\em red} for the Nexus 5x and {\em black}
for the OnePlus One.

The goal of using these two phones is to understand how larger form
factors, which provide more viewable space, may affect the attackers
ability to shoulder surf ({\bf H6}). There are also side effects for a larger
display that we did not anticipate. For example, in
Figure~\ref{fig:phones}, with the larger OnePlus One phone, we
experienced more glare on the screen as it was a bit more unwieldy. Being larger in the hand, the OnePlus phone moved more during PIN/Pattern entry, particularly one-handed, which caused more opportunities for glare.

\paragraph{Hand positions}
We investigated two different phone-grips (or hand positions) for
authentication entry. Figure~\ref{fig:phones} shows the grips.  The images on the top-left and bottom-left show a single handed grip being used, where the thumb is used to enter the authentication. The images at the top-right and bottom-right, the grip is a two handed grip, where the
user holds the phone in their left hand and enters the authentication sequence using
the index-finger of their right hand. These are both common grip
settings for mobile devices~\cite{eardley2017grip}. We focus
exclusively on right handed entry modes to reduce the complexity of
our experiment.  In charts and tables, we describe these two hand
positions as {\em thumb} for the single handed grip with thumb entry,
and {\em index} for the two handed grip with index finger entry.

We applied these two conditions because we hypothesized ({\bf H7})
that visual obstructions may impact the vulnerability to shoulder
surfing. For example, using an index finger provides the least
obstructed view, compared to using the thumb, but it also may increase
point-of-view obfuscation where it may appear that contact is being
made with the phone, when it is only an illusion due to the the angle of observation.

\begin{figure}
\small
\centering
\includegraphics[width=0.45\linewidth]{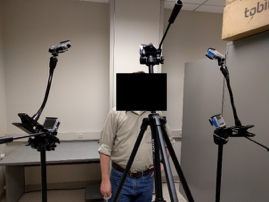}
\caption{GoPro Camera Array: lower cameras are {\em near}, higher cameras are {\em far}, and the middle camera is {\em top}}
\label{fig:array} 
\end{figure}
\begin{table}
\small
\centering
    \begin{tabular}{M{.75in} | M{.6in} | M{1in}}
      \textbf{Angle Name} & \textbf{Visual} & \textbf{Description}\\
      \hline
      Near Left (nl) & \includegraphics[width=0.4\marginparwidth]{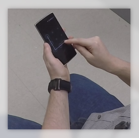} & Over target's left shoulder at a height of 5'\\
\hline
      Far Left (fl) & \includegraphics[width=0.4\marginparwidth]{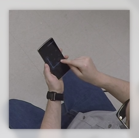} &Over target's left shoulder at a height of 6'\\
\hline
      Top (t)& \includegraphics[width=0.4\marginparwidth]{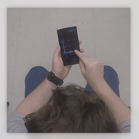} & Over target's head at a height of 6'\\
      \hline
      Near Right (nr) &  \includegraphics[width=0.4\marginparwidth]{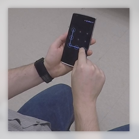} &Over target's right shoulder at a height of 5'\\
      \hline
      Far Right (fr)& \includegraphics[width=0.4\marginparwidth]{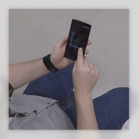} &Over target's right shoulder at a height of 6'\\
  \end{tabular}
  \caption{Camera Angles}
  \label{tab:angles}
\end{table}

\paragraph{Angles of recording}
We used a camera array to simultaneously record each authentication
(e.g., one phone type, one hand position type, one authentication
input) from multiple angles. The camera array is shown in
Figure~\ref{fig:array}.  The target user, who is seated for the study, is subject of observations from five angles in the camera array to simulate different vantage points. Outlined in Table~\ref{tab:angles}, the
angles are, from each side {\em left} and {\em right} with a {\em far}
and {\em near} angle. We also had a {\em top} angle with a
vantage immediate overhead of the target user. In charts and tables,
we shorthand these angles as: {\em nl} for near left, {\em fl} for
far left, {\em t} for top, {\em nr} for near right, and {\em fr} for
far right.

We hypothesized that there may exist settings of observations that
both hinder and enhance the attackers ability to shoulder surf ({\bf
  H4}). For example, observations from one side over the other (e.g.,
left v. right) may provide more or less obstructed views, aiding or
hindering shoulder surfing.


\begin{figure}
  \centering
  \includegraphics[width=0.65\linewidth]{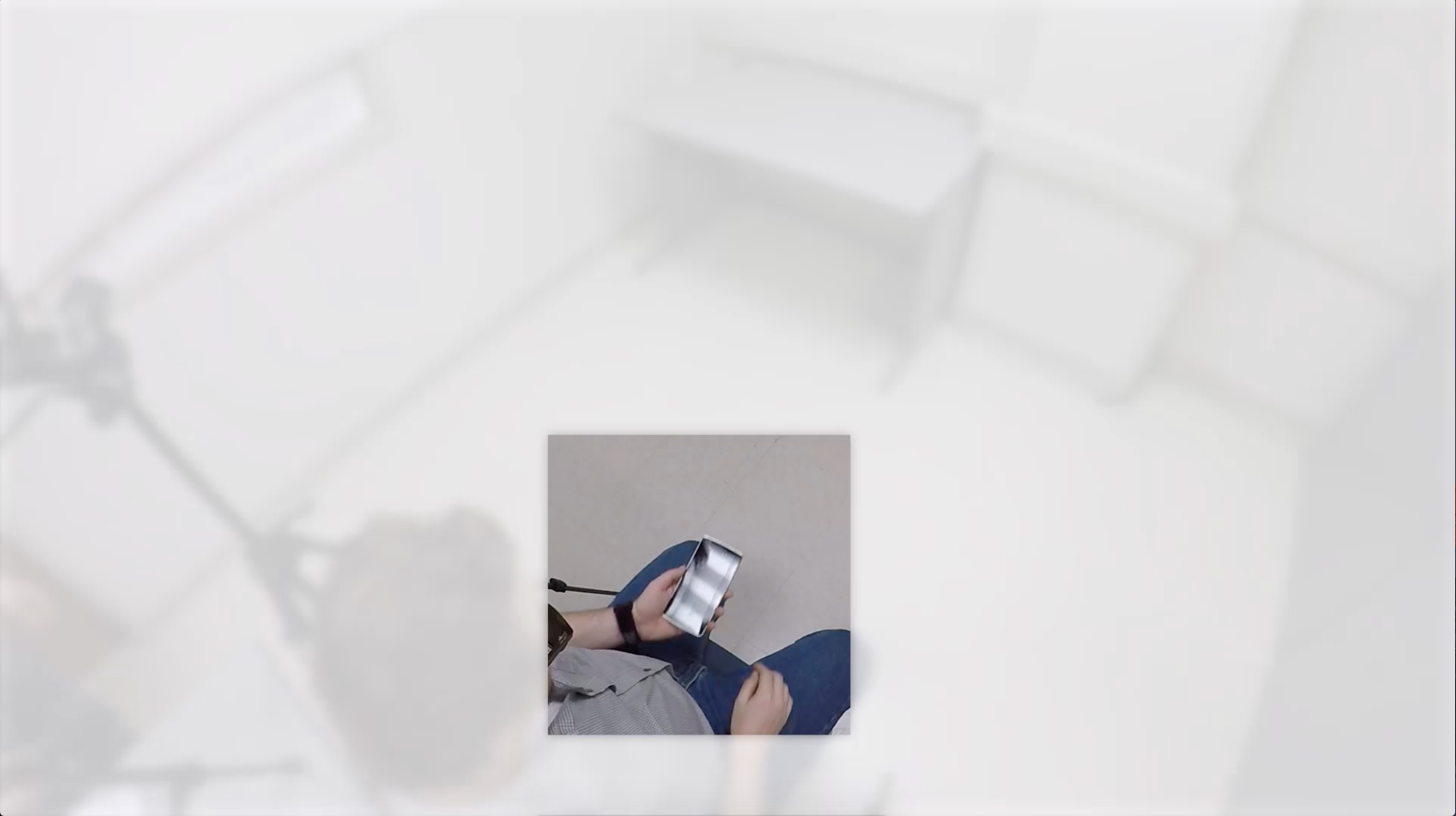}
  \caption{Full screen video with ``focus zone'' to highlight where to
    view the authentication and to remove distractions from other
    factors in the recording space.}
  \label{fig:focus}
\end{figure}

\paragraph{Editing Videos}
During video recordings, we attempted to make each authentication occur over a consistent length of time with a consistent hand motion.
We further attempted to
remove any distractions from the observation area so that participants
can focus directly on the task of shoulder surfing. Each video
recording, is about 3-5 seconds in length, but this creates a tracking
challenge for the participant who needs to quickly determine where to
look in a video (occurring from different angles each time) to do the
observation. To alleviate this burden, we edited the videos by placing
a ``focus zone'' in the video. See Figure~\ref{fig:focus} for a visual
of this editing. Except for the authentication area, the remainder of
the screen is set with a transparent gray so that the participant can
quickly determine where to focus their visual attention for the observation task.

\subsection{Authentication Settings}
As previously mentioned, we aim to analyze two different
authentication settings, PINs and Android graphical pattern
unlock. Within the Android pattern settings, we also consider settings
where the tracing lines are either displayed or not displayed. Recent work has suggested
that tracing lines should {\em not} be displayed for improved
security~\cite{vzw2015easy}. For each authentication setting, we
have chosen a set of 10 representative PINs and patterns that have spatial
shifting properties and visual, complexity properties, such as crosses.

In the remainder of this section, we outline how that selection was
performed and justify the properties used during
selection. Additionally, we describe the application used for
performing input and how it was designed to fairly compare the two
authentication types.

\begin{table}[t]
\centering
\small
  \begin{tabular}{ r | c  || r | c}
    PINs & properties & Patterns & properties \\
    \hline
1328 & up/non-adj              & 0145 & up\\
1955 & neutral/non-adj/repeats & 1346 & left \\
5962 & right                   & 3157 & neutral \\
6702 & down/kmove/cross        & 4572 & right/cross  \\
7272 & left/kmoves/repeats     & 6745 & down \\
\hline
153525 & up/repeat                & 014763 & left/cross \\
159428 & neutral/cross/non-adj    & 136785 & down \\
366792 & right/repeat/kmove/cross & 642580 & neutral/cross  \\
441791 & left/kmove/repeat        & 743521 & up/non-adj \\
458090 & down/repeat              & 841257 & right/kmove/cross \\
  \end{tabular}
  \caption{PINs and patterns used in experiments. See Appendix~\ref{fig:patterns} and~\ref{fig:pins} for visuals of the authentication.}
\label{tab:authentication}
\vspace{-.2in}
\end{table}

\paragraph{Pattern Selections} 
The patterns used in our experiment are shown in
Table~\ref{tab:authentication} (graphical representations are presented in the Appendix). These patterns were culled from a set of
self-reported patterns collected through an online
study~\cite{aviv2015isbigger}, and provided to us for analysis and
use. From these patterns, we identified five 4-length patterns and five
6-length patterns that exhibited a broad set of representative features. 

To determine which features to consider, we hypothesized that there may be locations in the grid space that increase or decrease the effectiveness of the attack ({\bf H5}), as well as complexity features of patterns~\cite{andriotis2013pilot,andriotis2014complexity}. We were guided by related work~\cite{aviv2014understanding,
  vonZezschwitz2016quant} in choosing the features, for both spatial aspects and complexity properties.
\begin{itemize}
\item {\em up} shifted: The contact points of the pattern are in the upper part of the grid space, such as pattern 0145 and 743521.
\item {\em down} shifted: The contact points of the pattern are in the
  lower part of the grid space, such as pattern 6745 and 136785.
\item {\em left} shifted: The contact points of the pattern are in the
  left portion of the grid space, such as pattern 1346 and 014673.
\item {\em right} shifted: The contact points of the pattern are in the right portion of the grid space, such as pattern 4572 and 841257.
\item {\em neutral}: The contact points are evenly distributed in the grid space. 
\item {\em non-adjacency} (or non-adj): Two, non-adjacent contact points are used,
  such as pattern 743521, where contact point 3 and 5 are non-adjacent
  and are connected because contact point 4 was contacted prior.
\item {\em knight move} (or kmove): Two contact points are connected
  over two and down one (or in any symmetry), like a knight moves in
  chess, like in pattern 4572. 
\item {\em cross}: The sequence of contacts crosses over itself, such
  as pattern 014673 and 841257 have a perfect `X', but more obtuse
  crosses also exist, such as in pattern 4572 due to a knight move.
\end{itemize}

\paragraph{PIN selections}
In order to select PINs, we followed related work in analyzing digit
sequences in password datasets~\cite{FC:BonPreAnd12}. Using the
RockYou dataset\footnote{Originating from a debunk music sharing web site, the
  RockYou dataset was leaked in 2009 and contains over 32 million
  passwords commonly used by researchers~\cite{CCS:WACS10}.},  we extracted 4- and 6-length digit sequences that exhibited similar
properties to that of the pattern dataset. The idea being that these
digit sequences are likely to be reused as PINs if they appear in
passwords.

Matching the PINs to the exact features in the patterns is not perfect,
as not all digit sequences found in patterns exist within the RockYou
dataset, and further, we wish to include all 10 digits (patterns only
use 9 contact points). PINs also have a feature that patterns cannot
have, repeated digits, so we wish to include PINs with this property,
either a single digit or multiple repeated digits. The final set of PINs
selected are available in Table~\ref{tab:authentication}, and a visual is provided in Appendix~\ref{fig:pins}.

\paragraph{Authentication Applications}
Another important factor to consider is the applications used for
entering the authentication. Critically, the size of each application
should be the same and have similar visual properties, so as not to
advantage one over the other for shoulder surfing.  To this end, we
designed two Javascript applications using HTML5 that ran in the
Android Chrome browser, setup as a home screen link to simulate a
standalone application. Each application mimicked the input used on
the device, following the same rules. During the survey, the same
applications would be used as embedded Javascript in their browser for
the participants to recreate the authentication observed.

For patterns with-line feedback, after the target user completed the
application, there would be a brief, 200 ms pause before the screen would
go to blank/black screen. This is to simulate the unlock process on the phone. A
similar action occurs for patterns without-line feedback,
however, no tracing lines or circled contact points would be seen.

For PINs, the input text area would show the number that was pressed,
but would fade to an asterisk after one second or after the next
number was pressed, similar to how unlock authentication works on
smartphones. Only after pressing ``ok'' would the screen go blank,
simulating an unlock.

\subsection{Survey Protocol}
We designed a protocol around the video recordings by which
participants would be assigned a randomized group, receiving training
relevant to that group, and then attempt to shoulder surf 10
authentications based on observing videos under different
settings. The survey was designed as a web application using a
combination of PHP, Javascript, and a MySQL backend. The survey was
posted on Amazon Mechanical Turk and participants were also recruited
locally at our institution to ensure consistency.

The survey protocol proceeded as follows:
\begin{enumerate}
  \item Informed Consent
  \item Demographic and Background Information
  \item Training
  \item Observations and Recreation
  \item Attention Check and Submission
\end{enumerate}
In the remainder of this section we outline each of these survey
segments in detail, as well as the randomization and recruitment
process.

\paragraph{Informed Consent and Preliminary Instructions}
This survey was approved by our institutional oversight board (IRB),
and so we require participants to provide informed consent. For online
participants, this was done digitally, and for in-person participants,
it was done in a traditional manner, following a script.  

The informed consent also provided participants with an overview of the
experiment, its goals, and initial instructions. For example, it
informed participants that they were participating in a research
project about shoulder surfing, as well as directions about the
procedures:
\begin{quote}
  \small \em
  The survey will request that you maximize the browser window on your
  screen. You are not permitted to record the survey or any of its
  content. The use of pen and paper to write anything down is also
  strictly prohibited. The survey will request that you watch several
  videos of a user authenticate into a mobile device. You are to watch
  the video and attempt to recreate the PIN or pattern you viewed
  being entered.
\end{quote}

\paragraph{Demographic and Background Information}
Following acknowledgment of the informed consent, we ask a series of
demographic questions. Including:
\begin{itemize}
  \item Gender (Male, Female, Prefer not to answer)
  \item Age (drop down box, 18-100)
  \item Eye Sight (Normal, Corrected with glasses/contacts, Deficient and not corrected)
  \item Ability with modern cell phones (None, Below Average, Average, Above Average, Professional)
\end{itemize}
Additionally, we recorded the screen size of the browser, in pixels,
to test if participants were following directions as well to get a sense
of the different viewing scenarios.

\begin{figure}
\centering
\includegraphics[width=0.7\linewidth]{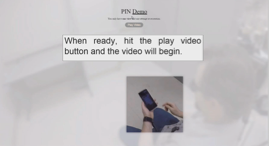}
\caption{Frame of tutorial video}
\label{fig:tutorial} 
\end{figure}

\paragraph{Between Treatment Randomization and Training}
At this point in the survey, we randomize the treatments as the
remainder portion is dependent on that randomization. We initially
randomize into 12 between subject treatment groups:
\begin{itemize}
\item Authentication Type: PIN, patterns with feedback lines, or
  patterns without feedback lines
\item Hand Position: index or thumb
\item Phone Type: either the red Nexus 5 or the black OnePlus One
\end{itemize}
Based on the selection, we prepared three training videos that
explained the procedure further specifically for each authentication
type, and then 12 sample test videos that participants can use to
practice shoulder surfing. The test video used the same conditions as
the selected treatment, but with a sample PIN (1234) or Pattern
(0123). The training video shows the participant how the observation and recall would
proceed (a screen-shot of the video is in Figure~\ref{fig:tutorial}), and once the video completes, test runs are performed
using the sample PIN or pattern. Participants are allowed to repeat
this training video and test runs as many times as needed before
continuing to the main portion of the survey.

\begin{table}[t]
\centering
\small
\begin{tabular}{c | c | c}
\textbf{Conditions} & \textbf{Views}  & \textbf{Attempts}\\
\hline
 A & One & One\\
 B & One & Two\\
 C & Two & One\\
 D & Two (different angles) & One\\
 E & Two (different angles) & Two\\
\end{tabular}
\caption{Five different conditions for each authentication}
\label{tab:treatments}
\end{table}

\begin{table}[t]
  \centering
  \resizebox{\linewidth}{!}{
  \begin{tabular}{ c r | c c | l || c c c | l}
    & & \multicolumn{3}{c||}{{\em In-person}} & \multicolumn{3}{c|}{{\em Online (MTurk)}}\\
    & & Male & Female & Total & Male & Female & Non-Spec. & Total\\
    \hline
    \parbox[t]{2mm}{\multirow{6}{*}{\rotatebox[origin=c]{90}{{\em Age}}}}%
    & 18-24 & 68 &23 & 91& 103 & 78 &  & 181 \\
    & 25-34 &  &  & & 304 & 221 & 6 & 531 \\
    & 35-44 &  &  & & 142 & 120 & 1 & 263 \\
    & 45-54 &  &  & & 47 & 87 &    & 134\\
    &54-64 &  &  & & 21 & 27 &   & 48 \\
    &  65+ &  &  & & 7  & 8 & 1   & 16 \\
    \hline 
    \parbox[t]{2mm}{\multirow{3}{*}{\rotatebox[origin=c]{90}{{\em Sight}}}}%
    &Deficient &  &  &  & 7 &  3 &   & 10\\
    &Corrected & 12 & 9 & 21 &  225 & 252 & 3 & 480 \\
    &Normal & 56 & 14 & 70 &  392 & 286 & 5 & 684\\
    \hline
    \parbox[t]{2mm}{\multirow{4}{*}{\rotatebox[origin=c]{90}{{\em Skill}}}}%
    &Below         &  &    &  &  17 & 11 & 9 & 37 \\ 
    &Below Average & 22 & 10 & 32&  134 & 204 & 6 & 344 \\ 
    &Above Average & 38 & 12 & 50&  344 & 277 & 1 & 622 \\ 
    &Professional  &  8 & 1  & 9&  129 & 49 & 1 &  179\\ 
    \hline
    \parbox[t]{2mm}{\multirow{4}{*}{\rotatebox[origin=c]{90}{{\em Resolution}}}}%
          & $<1300$ &  &&&  117 & 119 & 4 &  240\\
    & 1300-1500 &  &&&  189 & 250 & 1 & 440\\
    & 1500-1800 & &&&  122 & 85 & 2 & 209 \\
    & $>1800$ & 68 & 23 &&  196 & 87 & 1 & 284\\
    \hline 
    & & 68  & 23 & {\bf 91} &  624 & 541 & 8 & {\bf 1173}\\
  \end{tabular}}
\caption{Demographic Information. The {\em resolution} refers to the width of the screen resolution, in pixels.}

  \label{tab:demo}
\end{table}


\begin{table*}[t]
  \small
  \centering
  \begin{tabular}{ r | c c |  c c | c || c c |  c c | c }
    & \multicolumn{4}{c| }{\em Online (MTurk)} & & \multicolumn{4}{c|}{\em In-person} \\
    & \multicolumn{2}{c|}{\em Single View} & \multicolumn{2}{c|}{\em Multi View}  & & \multicolumn{2}{c|}{\em Single View} & \multicolumn{2}{c|}{\em Multi View}\\

    & 4-length & 6-length  & 4-length & 6-length & {\bf total}   & 4-length & 6-length  & 4-length & 6-length & {\bf total}\\
    \hline
     PIN  & 34.92\% & 10.86\% & 56.72\% & 26.53\% & 32.25\%  & 52.63\% & 20.75\% & 76.62\% & 59.32\% & 46.02\% \\
    NPAT  & 51.03\% & 35.28\% & 71.27\% & 52.10\% &  52.28\% & 72.29\% & 62.31\% & 84.31\% & 95.38\% &  72.92\% \\
    PAT   & 80.90\% & 64.20\% & 88.07\% & 79.85\% & 78.27\%    & 94.5\% & 86.74\% & 98.61\% & 83.53\% & 92.42\% \\
    \hline
    {\bf total}      & 55.73\% & 37.28\% & 72.15\%& 52.81\% &  & 73.70\% & 56.70\% & 86.5\%& 80.38\% &  \\
      & \multicolumn{2}{c|}{46.45\%} & \multicolumn{2}{c|}{62.53\%} & 54.54\%       & \multicolumn{2}{c|}{65.11\%} & \multicolumn{2}{c|}{83.37\%} & 70.81\%
  \end{tabular}
  \caption{Single- vs. multi-view for authentication types broken up based on online and in-person participants. NPAT is pattern {\em without} feedback lines and PAT is with feedback lines. Comparing single vs. multi-view, in all categories, was statistically significant, as well as in-person vs online.}
\label{tab:views}
\end{table*}


\paragraph{Within Treatment Randomization for Observation and Recall}
At this point, a participant has been assigned an authentication type,
phone type, and hand position. There is now a large set of videos from
multiple angles for each of the authentications, but it is not
feasible (nor desirable) to display every video to each
participant. Instead, we proceed with a within-group randomization to
display a subset of those videos under different settings that will
support testing hypothesis {\bf H2}, {\bf H3}, and {\bf H4}.

The first stage of randomization is to randomize the order of the
authentication that will be displayed. That is, each participant will
observe all 10 of the authentications in their selected authentication
type, either 10 PINs or 10 patterns, but the order of those must be
randomized to handle training effects where the participants become
better at the task as time goes on. Once the order is randomized, for
each authentication, we then randomize and counterbalance a set of conditions
regarding how many views and attempts a participant gets to make, as
outline in Table~\ref{tab:treatments}.

The ``views'' refer to how many times a participant gets to view
an observation video. For conditions A-C, a random angle is selected,
and the participant either gets a single view of that authentication
(A,B), or two views from the same angle (C). For conditions D and E,
participants get a random first angle selection, and then are assigned
a second angle on the opposite side (e.g., first angle is a left side,
second angle is a ride side). If the top angle was selected, then a
random second angle is used. 

The second part of each condition is the number of attempts. After
viewing the video, the participant can make either one attempt to
recreate the authentication or two attempts.

Prior to each video observation, we informed the participant if they
were going to view one or two videos and if they would have one or two
attempts.

\paragraph{Submission and Attention Tests}
Following the survey, we ask participants to report if they used
additional aids, such as pen and paper, in helping them complete the
procedure. This acts as both an attention test and a guide for
including or excluding results. It also allows us to exclude
participants who failed to follow directions. We did not have anyone
report that they ``cheated.''

\subsection{Recruitment}
We recruited locally at our institution, and online via Amazon
Mechanical Turk. The goal of using both recruitment methods is that
for the institutionally recruited participants, we can control the
settings, and so we wished to compare these results to those collected
online for consistency (see Figure~\ref{fig:reso}). Inconsistent
results would suggest that online participants were not taking the
survey faithfully. We observed consistent results when comparing
similar demographic groups with similar screen resolutions, as
described later, suggesting that participants online took the survey
in the intended ways. Although, there was some degradation of
performance, which may be accounted for by an observation bias or
the Hawthorne effect; local participants, being observed, were more likely
to try and perform the task well to appease their observers.

In total, we recruited 91 participants locally at our institution, and
1173 online participants. The demographic information is available in
Table~\ref{tab:demo}. The material used in recruitment mimicked that
of the informed consent. The text used in posting the task to Amazon
Mechanical Turk is provided in Appendix~\ref{app:advert}.

\subsection{Realism and Limitations}
We acknowledge that our experimental methodology has a number of
limitations. Foremost, we had to reduce the set of authentication
tokens to a reasonable size, namely 10, so that we could maintain a
reasonable survey length with a reasonable recruitment size. We
attempted to mitigate this effect by choosing real authentications, as
collected in other datasets, that would be representative of authentication
choice broadly. We further did not include text-based passwords, which
can form an unlock authentication, as we were unable to develop a
protocol to fairly compare to the other authenticators.

We were additionally limited in terms of the observation settings. Our
online participants may have used screens that were bigger or smaller than we
anticipated. We attempt to manage this limitation by recording the
screen size, and, as we will show in this paper, there was an impact on performance
with respect to screen size.  However, general trend lines remain the same,
when we compare the online data to that collected in-person.

\subsection{Ethical Considerations}
This protocol was reviewed and approved by our institution review
board to ensure that participants were treated fairly, such as
providing informed consent and an option to opt-out. The survey itself
does not elicit ethical challenges as participants are not performing
actions that increase the risk to others or themselves in regard to
shoulder surfing.  It could be argued these participants may be more
aware of the risks associated with these attacks after having
participated. The identity of the target victim was protected from
participants via obfuscation. Finally, the analysis does not include
identifiable information about participants.



\begin{table*}[t]
\begin{tabular}{r | c c l | c c l | c c l }
  & \multicolumn{3}{c|}{\em Hand Position} &  \multicolumn{3}{c|}{\em Phone Type} &   \multicolumn{3}{c}{\em Input Attempts}\\
  & Index & Thumb & $G$-test & Red & Black & $G$-test & One & Two & $G$-test\\
\hline
PIN  & 32.74\% & 32.22\%&  $p=0.68$ &30.84\%  & 34.04\%  & $p<0.05$ & 35.51\% & 30.43\% & $p<0.00005$\\
NPAT & 53.82\% & 50.75\% & $p<0.05$ &76.22\% & 80.23\%  & $p<0.05$ &56.56\% & 49.38\% & $p<1\times 10^{-7}$\\
PAT  & 79.93\% & 76.69\%& $p<0.05$ & 53.40\%  & 50.97\%  & $p<0.0005$ & 79.90\% & 77.16\% & $p<1 \times 10^{-9}$\\
\hline
{\bf total} & 55.30\%& 53.78\% & $p<0.05$ & 52.92\% & 56.06\% & $p<0.0005$ & 57.46\%  & 52.56\% & $p<0.005$\\ 
\end{tabular}
\caption{Hand position, phone type, input attempts, and observation angle impact for online participants. For hand position and phone type, single and multi-view treatments are considered. }
\label{tab:hand}
\label{tab:phone}
\label{tab:input}
\vspace{-.2in}
\end{table*}


\begin{figure}[t]
  \includegraphics[width=\linewidth]{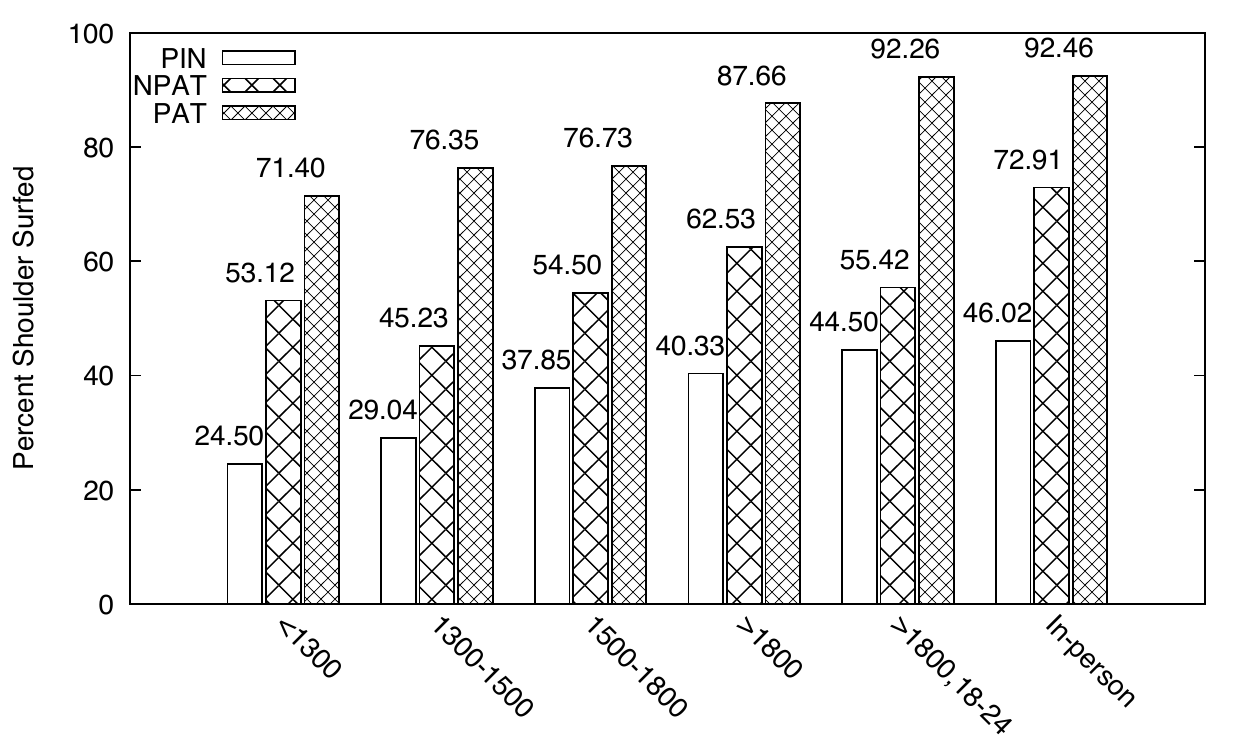}
\vspace{-.2in}
  \caption{Screen width resolution comparison. In-person subjects used
    a screen with resolution 990x1840 and are in the age range of
    18-24. There is no significant difference ($G$-Test) between PIN
    and PAT, there remains a significant difference for NPAT when
    comparing in-person samples to on line ones with similar
    resolutions and demographics. }
  \label{fig:reso}
\vspace{-.2in}
\end{figure}

\section{Results}
In this section, we describe the results of the survey by addressing each of the hypotheses outlined earlier. We also provide other insights as available, particularly related to the realism of the experiment. As
we move through the results, it is important to note that in some conditions (C and E) participants had multiple attempts to recreate the observed authentication, which we study in more detail later.  Unless otherwise noted, we consider a successful attack if the participants accurately recalled and entered the authentication sequence within either of the attempts.

For statistical testing, our data is categorical and binary, as in a
participant either correctly recalled and entered an authentication
sequence or did not. As such, in two way comparisons of attack rates,
we applied Fisher's Exact Test (or $G$-test) to test significance, and
$\chi^2$ test for comparing for multi-factor analysis. Additionally,
we perform a L1-penalty logistic regression analysis to determine the
impact (or lack thereof) of all settings of the experiment. A
significance level of $p<0.05$ is used. Finally, unless otherwise
stated, each of the tables, when a percentage is displayed, this
refers to the rate in which an authentication was successfully
recreated in that setting, a so called success or attack rate.

\paragraph{Realism of online results} 
An important question to consider is if online results are consistent
with those collected in-person. As evident in Table~\ref{tab:views},
there is a significant performance improvement for those in-person
participants ($p < 0.005$, using $\chi^2$). Investigating this
phenomenon further, we broke down the participants based on the width
of the screen resolution used while taking the survey, which is a good
approximation for the size of their viewing area. These results are
presented in Figure~\ref{fig:reso}, and one can clearly see that as
the resolution width increases, so does the performance. As we
controlled our in-person computing setup, we know precisely the screen
resolution of 990x1840, and further, our in-person participants (being
undergraduates) are between the ages of 18-24. When isolating this
demographic group, we find no statistical differences between the PIN
and PAT results, with remaining difference for patterns without
traceback lines (NPAT). This difference is likely the result of an
observation bias, by which having the researchers present led the
participants to want to perform the task ``better.'' As such, we find
that these results suggest that online participants likely took the
survey in the intended manner, and variations in screen size (and
other factors) probably realistically mimic the realities of shoulder surfing
in the wild. The remaining results focus solely on the online dataset.

\begin{table}[t]
  \resizebox{\linewidth}{!}{
\small
  \begin{tabular}{r | l l l l l }
\multicolumn{6}{c}{\em Observation Angle}\\
    &far-right & far-left & near-right & near-left & top\\
    & \includegraphics[width=0.4\marginparwidth]{figures/far_right.png} & \includegraphics[width=0.4\marginparwidth]{figures/far_left.png} & \includegraphics[width=0.4\marginparwidth]{figures/near_right.png} & \includegraphics[width=0.4\marginparwidth]{figures/near_left.png} &
\includegraphics[width=0.4\marginparwidth]{figures/top.png}\\
\hline
PIN & 20.39\%** & 18.88\%**& 21.77\%**& 32.46\%**&22.09\%**\\
NPAT  & 46.09\%& 35.62\%**& 46.57\%& 41.89\%**& 45.14\%*\\
PAT  & 70.31\%**& 68.23\%** & 73.47\%** & 71.50\%**& 79.08\%**\\ 
\hline
{\bf total} &  45.77\%** & 41.03\%** & 47.43\%*& 49.15\%& 49.18\% \\ 
  \end{tabular}}
\caption{Impact of observation angle on shoulder surfing. Single-view treatments and only multi-view treatments of the {\em same} angle are considered (see Table~\ref{tab:views} for single- vs. multi-view). Using $\chi^2$ testing, $*$ indicates $p<0.05$, $**$ indicates $p<0.005$. }
\label{tab:angle}
\vspace{-.2in}
\end{table}

\paragraph{H1: authentication type}
We applied both Fisher's exact test and $\chi^2$ test to the data in
Table~\ref{tab:views} and found all comparisons between authentication
type to be significant. Focusing on the online results, we find that
the authentication plays a significant role. PINs proved the most
elusive in all settings, with combined performance of 32.25\% attack
rate. Patterns with traceback lines was the worst performing, 78.27\%
attack rate across all settings. Removing traceback lines improved
results to 58.28\%, confirming prior work on this
topic~\cite{vzw2015easy}. \textbf{As such, we {\em accept} the hypothesis that
the authentication type impacts shoulder surfing vulnerability.}

\paragraph{H2: repeated observations}
Using the results in Table~\ref{tab:views}, we find that there are
significant differences between the single-view and multi-view
settings. Looking at both online and in-person results, participants
are about 1.3x-1.4x more likely to correctly attack an authentication if
allowed multiple views of the authentication. Later, as we compare all
the features, we find that multiple views, in particular, play an
outsized role in the vulnerability of authentication to shoulder
surfing. \textbf{As such, we {\em accept} the hypothesis that multiple
observations impact shoulder surfing vulnerability.}

\paragraph{H3: multiple input attempts}
Recall that we applied a within-group randomization by which some
participants on some authentication were provided two attempts to
input the authentication. The procedure of the survey informed them of
this fact, so participants were aware, prior to viewing the video,
that they would have multiple recreation
attempts. Table~\ref{tab:input} shows these results on the {\em right}
column.

Surprisingly, multiple attempts decrease performance, in all cases. We
believe this is because participants, knowing they would have multiple
attempts, attempted to ``game'' the process in a way that actually led
them to get the pattern wrong in both attempt cases. For example, they
would pay attention less well. \textbf{We {\em accept} the hypothesis
  that multiple input attempts affect shoulder surfing}, but it
decreases performance, unexpectedly. From this result, researchers
should consider for similar experiments to either not informing
participants how many attempts at recreation, or force participants
into a regime of single attempts, requiring more attention during that
single attempt.

\paragraph{H4: observation angle}
Table~\ref{tab:angle} presents the results comparing performance for
the different observation angles. As we wished to isolate the angle,
we only consider treatments where a single observation angle was
used. We used a $\chi^2$-test to determine significance factors in these
scenarios, indicated with *'s in the table. In nearly all cases,
within each authentication type, we found that there are significant
impacts based on the angle of observation. When performing comparisons in
total, we find that the far-right, far-left and near-right angles
showed the most significant impact. The far-left angle, in particular,
was the most challenging angle, and we believe that this angle provide
some obfuscation of when screen touching occurred, making it harder
for participant to cleanly determine the location of touch
events. \textbf{As such, we {\em accept} the hypothesis that the
  observation angle affects shoulder surfing.}

\begin{table}[t]
\centering
\small
  \begin{tabular}{r | l  || r | l  || r | l }
    \multicolumn{2}{c||}{\em PIN} &  \multicolumn{2}{c||}{\em NPAT} &   \multicolumn{2}{c}{\em PAT} \\ 
    \hline
1328 & 50.36\%  & 0145 & 81.10\%** & 0145 & 92.28\%**\\
1955 & 47.89\%  & 1346 & 43.18\%*  & 1346 & 82.96\%** \\
5962 & 36.10\%**  & 3157 & 60.26\%**  & 3157 & 87.09\%**\\
6702 & 34.64\%**  & 4572 & 47.90\%  & 4572 & 74.12\%** \\
7272 & 60.53\%**  & 6745 & 73.70\%**  & 6745 & 86.51\%** \\
\hline
153525 & 15.40\%** & 014763 & 53.44\% & 014763 & 84.01\%** \\
159428 & 17.17\%** & 136785 & 41.81\%* & 136785 & 73.76\%** \\
366792 & 19.57\%** & 642580 & 46.49\%  & 642580 & 74.03\%** \\
441791 & 20.66\%**  & 743521 & 37.90\%** & 743521 & 53.79\% \\
458090 & 21.83\%**  & 841257 & 37.78\%** & 841257 & 73.93\%** \\
  \end{tabular}
  \caption{Individual authentication attack rate. Significance tested using $\chi^2$ within authentication  of the same length, * indicating $p<0.05$ and ** indication $p<0.005$, or much less than.}
\label{tab:auth}
\vspace{-.2in}
\end{table}

\paragraph{H5: properties of authentication}
We first consider the length of the authentication, the results of
which are displayed in Table~\ref{tab:views}. The length has a large
impact. In most cases, it decreased the rate of shoulder surfing by
nearly 50\%. While length is far from a perfect approximation for
security, it's clear that longer authentication will improve security
from observation attack. We further breakdown the vulnerability of the
individual authentications in Table~\ref{tab:auth}. Many of the
authentications vary from an expected uniform attack rate, as observed by using a $\chi^2$ test within
each authentication length. However, there does not appear to be a
direct pattern related to the individual spatial properties of the
authentication, additional analysis with more authentication types
would be needed to draw strong conclusions regarding these
features. \textbf{As such, we {\em partially accept} the hypothesis
  that the properties of authentication impact shoulder surfing,}
while features such as authentication length play a large role, the
impact of other features is inconclusive.

\paragraph{H6: phone size}
Table~\ref{tab:hand} shows the result of comparing the two phones in
the study. Recall that the Red phone refers to the 5'' display Nexus
5, and the Black phone refers to the 5.5'' display of OnePlus
One. Across all conditions, we find that there is a significant
difference in shoulder surfing  between the two phones. In most
cases, the larger Black phone provides less security, except for
patterns (PAT), where the smaller Red phone is more secure. After
reviewing the videos, we noticed that the larger Black phone
experiences more glare during this recording which could account for
the difference. Overall, it appears that the larger phones provide
less security for shoulder surfing, and \textbf{ we accept the
  hypothesis.}

\paragraph{H7: hand position}
Recall that we examined two different hand positions. One hand
position (or grip) had the victim use a single hand, entering the
authentication with his thumb. The second hand position was two
handed, holding the phone in the left hand entering the patterns with
the right index finger. Table~\ref{tab:hand} shows the results of
comparing these two conditions, {\em thumb} vs {\em index}. The
results for comparing PINs showed no significant difference; however
there was significant, but small, differences between pattern entry for the different hand positions,
as well as a small significant difference overall. While there is a
difference, the impact factor is challenged, so \textbf{we {\em
    reject} the hypothesis that hand position impacts shoulder
  surfing.} These results suggest that researchers can allow for any
normal hand position without greatly impacting the results; however,
using an index finger provides a more direct view, as opposed to the
one-handed thumb blocking portions of the screen) and likely improves
results, nominally.

\paragraph{Comparison across features}
Finally, we wish to understand how the combination of the features
impact the results, asking the question, are there a set of ideal
conditions or non-ideal conditions for shoulder surfing that can form
a set of baselines? To accomplish this, we performed a logistic
regression across all the features using L1-penalties such that
features that have small (or no) effect can have a coefficient of
zero. The results of an average of 100 runs of the regression (there
were many different minimums) are presented in Table~\ref{tab:log}.

The regression was set-up using a feature set of binary values, with a
one indicating the presence of the feature and zero otherwise. The
label on the feature was also binary, a zero indicating that shoulder
surfing attack failed and one indicating success. We trained over each trial
of the survey, and the resulting model was able to explain 68.7\% of
the data and was significant. 

We can further analyze the coefficients of the features which indicate
how much weight they provide to the prediction and also if they
increase or decrease the likelihood of shoulder surfing. Negative
values imply greater security to shoulder surfing, while positive
values indicate more vulnerability to shoulder surfing. As we are
using L1 penalty, some coefficients can reduce to zero.

Most surprisingly, the coefficient for NPAT (patterns without tracing
lines) is 0. This makes sense if you consider the fact that being a
PIN so greatly reducing the likelihood of shoulder surfing, while
patterns greatly increase the likelihood. The fact that it is a
pattern without lines is not predictive, in comparison to those other
two facts. Further, the highest coefficient is that of shoulder
surfing a pattern, followed by PINs (in the negative direction). This
further supports accepting hypothesis {\bf H1}.

Among the other coefficients, the length factor and having multiple views of the authentication play a large role in  shoulder surfing attack rates. The far-left angle proved to be
the most challenging for shoulder surfing, as identified
earlier while near-left and top were the most beneficial for shoulder
surfing.

Based on these results, we can now identify category of the best case
scenario for an attacker performing shoulder surfing: attacking a
pattern with tracing lines that is of length four when provided
multiple with views. Similarly, the worst case scenario is attacking a PIN
of length 6 from the far left when provided with just one view. These two
scenarios can provide a baseline to compare new systems that offer
protections to shoulder surfing, as well as help inform users of stronger
authentication choices.


\begin{table}[t]
\small
\centering
\begin{tabular}{r | c  c }
 Feature & Coefficient \\
\hline
  PIN & -0.90**\\
  PAT & 1.25** \\
  NPAT & 0.00 \\
\hline
  4-length & 0.42*\\
  6-length & -0.33*\\
\hline
  Thumb & -0.03 \\
  Index & 0.06 \\
\hline
  Red  & -0.03\\
  Black & 0.09\\
\end{tabular}
\hspace{.1in}
\begin{tabular}{r | c  c}
 Feature & Coefficient\\
\hline
  Single-View & -0.09\\
  Multi-View & 0.40*\\
\hline
  One-Input & 0.11*\\
  Two-Input &-0.10*\\
\hline
  far-left & -0.24* \\
  far-right & -0.06 \\
  near-left & 0.12*\\
  near-right & 0.00\\
  top & 0.14*\\
\end{tabular}
\caption{L1-penalty logistic regression using all features, the average of 100 runs of the regression. 68.7\% of the data is explained by the regression. The * indicate top ranked coefficients. The model is significant. }
\label{tab:log}
\vspace{-.3in}
\end{table}


\section{Conclusion}
We presented the results of a large scale, online study of shoulder
surfing for the most common unlock authentication, PINs, patterns with
tracing lines, and patterns without tracing lines. We find that PINs
are the most secure to shoulder surfing attacks, and while both types
of pattern input are poor, patterns without lines provides greater
security. The length of the input also has an impact; longer
authentication is more secure to shoulder surfing. Additionally, if
the attacker has multiple-views of the authentication, the attacker's
performance is greatly improved. 

Overall, the goal of this research is to work towards establishing
baselines for how current authentication performs against shoulder
surfing, as well as provide insight into settings of current
authentication that can protect users from shoulder surfing. Based on
our analysis, researchers should consider comparing their performance
of new systems to the most secure setting, namely using at least
6-digit PINs with just a single view, as well as to the least secure
setting of using a 4-length pattern with visible lines with multiple
views. Additionally, these results suggest, for users, that 6-digit
(or longer) PINs provide the best security from shoulder surfing.


\section{Acknowledgments}
We thank Courtney Tse for her assistance conducting user studies.  This research is funded by the National Security Agency and the Office of Naval Research (N00014-15-1-2776).

\bibliographystyle{ACM-Reference-Format}
\bibliography{crypto,abbrev1,ref}


\begin{thebibliography}{00}


\ifx \showCODEN    \undefined \def \showCODEN     #1{\unskip}     \fi
\ifx \showDOI      \undefined \def \showDOI       #1{#1}\fi
\ifx \showISBNx    \undefined \def \showISBNx     #1{\unskip}     \fi
\ifx \showISBNxiii \undefined \def \showISBNxiii  #1{\unskip}     \fi
\ifx \showISSN     \undefined \def \showISSN      #1{\unskip}     \fi
\ifx \showLCCN     \undefined \def \showLCCN      #1{\unskip}     \fi
\ifx \shownote     \undefined \def \shownote      #1{#1}          \fi
\ifx \showarticletitle \undefined \def \showarticletitle #1{#1}   \fi
\ifx \showURL      \undefined \def \showURL       {\relax}        \fi
\providecommand\bibfield[2]{#2}
\providecommand\bibinfo[2]{#2}
\providecommand\natexlab[1]{#1}
\providecommand\showeprint[2][]{arXiv:#2}

\bibitem[\protect\citeauthoryear{Ali, Aviv, and Kuber}{Ali
  et~al\mbox{.}}{2016}]%
        {ali2016developing}
\bibfield{author}{\bibinfo{person}{Abdullah Ali}, \bibinfo{person}{Adam~J
  Aviv}, {and} \bibinfo{person}{Ravi Kuber}.} \bibinfo{year}{2016}\natexlab{}.
\newblock \showarticletitle{Developing and evaluating a gestural and tactile
  mobile interface to support user authentication}.
\newblock \bibinfo{journal}{{\em IConference 2016 Proceedings\/}}
  (\bibinfo{year}{2016}).
\newblock


\bibitem[\protect\citeauthoryear{Andriotis, Tryfonas, and Oikonomou}{Andriotis
  et~al\mbox{.}}{2014}]%
        {andriotis2014complexity}
\bibfield{author}{\bibinfo{person}{Panagiotis Andriotis}, \bibinfo{person}{Theo
  Tryfonas}, {and} \bibinfo{person}{George Oikonomou}.}
  \bibinfo{year}{2014}\natexlab{}.
\newblock \showarticletitle{Complexity metrics and user strength perceptions of
  the pattern-lock graphical authentication method}.
\newblock In \bibinfo{booktitle}{{\em Human Aspects of Information Security,
  Privacy, and Trust}}. \bibinfo{publisher}{Springer},
  \bibinfo{pages}{115--126}.
\newblock


\bibitem[\protect\citeauthoryear{Andriotis, Tryfonas, Oikonomou, and
  Yildiz}{Andriotis et~al\mbox{.}}{2013}]%
        {andriotis2013pilot}
\bibfield{author}{\bibinfo{person}{Panagiotis Andriotis}, \bibinfo{person}{Theo
  Tryfonas}, \bibinfo{person}{George Oikonomou}, {and} \bibinfo{person}{Can
  Yildiz}.} \bibinfo{year}{2013}\natexlab{}.
\newblock \showarticletitle{A pilot study on the security of pattern
  screen-lock methods and soft side channel attacks}. In
  \bibinfo{booktitle}{{\em Proceedings of the sixth ACM conference on Security
  and privacy in wireless and mobile networks}} {\em
  (\bibinfo{series}{WiSec'13})}. \bibinfo{pages}{1--6}.
\newblock


\bibitem[\protect\citeauthoryear{Aviv, Budzitowski, and Kuber}{Aviv
  et~al\mbox{.}}{2015}]%
        {aviv2015isbigger}
\bibfield{author}{\bibinfo{person}{Adam~J. Aviv}, \bibinfo{person}{Devon
  Budzitowski}, {and} \bibinfo{person}{Ravi Kuber}.}
  \bibinfo{year}{2015}\natexlab{}.
\newblock \showarticletitle{Is Bigger Better? Comparing User-Generated
  Passwords on 3x3 vs. 4x4 Grid Sizes for Android's Pattern Unlock}. In
  \bibinfo{booktitle}{{\em Proceedings of the 31st Annual Computer Security
  Applications Conference}} {\em (\bibinfo{series}{ACSAC 2015})}.
  \bibinfo{publisher}{ACM}, \bibinfo{address}{New York, NY, USA},
  \bibinfo{pages}{301--310}.
\newblock
\showISBNx{978-1-4503-3682-6}
\showDOI{%
\url{https://doi.org/10.1145/2818000.2818014}}


\bibitem[\protect\citeauthoryear{Aviv and Fichter}{Aviv and Fichter}{2014}]%
        {aviv2014understanding}
\bibfield{author}{\bibinfo{person}{Adam~J. Aviv} {and} \bibinfo{person}{Dane
  Fichter}.} \bibinfo{year}{2014}\natexlab{}.
\newblock \showarticletitle{Understanding Visual Perceptions of Usability and
  Security of Android's Graphical Password Pattern}. In
  \bibinfo{booktitle}{{\em Proceedings of the 30th Annual Computer Security
  Applications Conference}} {\em (\bibinfo{series}{ACSAC '14})}.
  \bibinfo{publisher}{ACM}, \bibinfo{address}{New York, NY, USA},
  \bibinfo{pages}{286--295}.
\newblock
\showISBNx{978-1-4503-3005-3}
\showDOI{%
\url{https://doi.org/10.1145/2664243.2664253}}


\bibitem[\protect\citeauthoryear{Aviv, Gibson, Mossop, Blaze, and Smith}{Aviv
  et~al\mbox{.}}{2010}]%
        {aviv2010smudge}
\bibfield{author}{\bibinfo{person}{Adam~J Aviv}, \bibinfo{person}{Katherine
  Gibson}, \bibinfo{person}{Evan Mossop}, \bibinfo{person}{Matt Blaze}, {and}
  \bibinfo{person}{Jonathan~M Smith}.} \bibinfo{year}{2010}\natexlab{}.
\newblock \showarticletitle{Smudge Attacks on Smartphone Touch Screens.}. In
  \bibinfo{booktitle}{{\em Proceedings of the 2010 Workshop on Offensive
  Technology}} {\em (\bibinfo{series}{WOOT'10})}. \bibinfo{pages}{1--7}.
\newblock


\bibitem[\protect\citeauthoryear{Bhagavatula, Ur, Iacovino, Kywe, Cranor, and
  Savvides}{Bhagavatula et~al\mbox{.}}{2015}]%
        {bhagavatula2015biometric}
\bibfield{author}{\bibinfo{person}{Chandrasekhar Bhagavatula},
  \bibinfo{person}{Blase Ur}, \bibinfo{person}{Kevin Iacovino},
  \bibinfo{person}{Su~Mon Kywe}, \bibinfo{person}{Lorrie~Faith Cranor}, {and}
  \bibinfo{person}{Marios Savvides}.} \bibinfo{year}{2015}\natexlab{}.
\newblock \showarticletitle{Biometric Authentication on iPhone and Android:
  Usability, Perceptions, and Influences on Adoption}.
\newblock  (\bibinfo{year}{2015}).
\newblock


\bibitem[\protect\citeauthoryear{Bianchi and Oakley}{Bianchi and
  Oakley}{2014}]%
        {bianchi2014multiplexed}
\bibfield{author}{\bibinfo{person}{Andrea Bianchi} {and} \bibinfo{person}{Ian
  Oakley}.} \bibinfo{year}{2014}\natexlab{}.
\newblock \showarticletitle{Multiplexed input to protect against casual
  observers}. In \bibinfo{booktitle}{{\em Proceedings of HCI Korea}}. Hanbit
  Media, Inc., \bibinfo{pages}{7--11}.
\newblock


\bibitem[\protect\citeauthoryear{Bonneau, Preibusch, and Anderson}{Bonneau
  et~al\mbox{.}}{2012}]%
        {FC:BonPreAnd12}
\bibfield{author}{\bibinfo{person}{Joseph Bonneau}, \bibinfo{person}{S{\"o}ren
  Preibusch}, {and} \bibinfo{person}{Ross Anderson}.}
  \bibinfo{year}{2012}\natexlab{}.
\newblock \showarticletitle{A Birthday Present Every Eleven Wallets? {T}he
  Security of Customer-Chosen Banking {PINs}}. \bibinfo{pages}{25--40}.
\newblock


\bibitem[\protect\citeauthoryear{Cherapau, Muslukhov, Asanka, and
  Beznosov}{Cherapau et~al\mbox{.}}{2015}]%
        {cherapau2015impact}
\bibfield{author}{\bibinfo{person}{Ivan Cherapau}, \bibinfo{person}{Ildar
  Muslukhov}, \bibinfo{person}{Nalin Asanka}, {and} \bibinfo{person}{Konstantin
  Beznosov}.} \bibinfo{year}{2015}\natexlab{}.
\newblock \showarticletitle{On the impact of touch id on iphone passcodes}. In
  \bibinfo{booktitle}{{\em Eleventh Symposium On Usable Privacy and Security
  (SOUPS 2015)}}. \bibinfo{pages}{257--276}.
\newblock


\bibitem[\protect\citeauthoryear{De~Luca, Denzel, and Hussmann}{De~Luca
  et~al\mbox{.}}{2009}]%
        {deluca2009look}
\bibfield{author}{\bibinfo{person}{Alexander De~Luca}, \bibinfo{person}{Martin
  Denzel}, {and} \bibinfo{person}{Heinrich Hussmann}.}
  \bibinfo{year}{2009}\natexlab{}.
\newblock \showarticletitle{Look into My Eyes!: Can You Guess My Password?}. In
  \bibinfo{booktitle}{{\em Proceedings of the 5th Symposium on Usable Privacy
  and Security}} {\em (\bibinfo{series}{SOUPS '09})}. \bibinfo{publisher}{ACM},
  \bibinfo{address}{New York, NY, USA}, Article \bibinfo{articleno}{7},
  \bibinfo{numpages}{12}~pages.
\newblock
\showISBNx{978-1-60558-736-3}
\showDOI{%
\url{https://doi.org/10.1145/1572532.1572542}}


\bibitem[\protect\citeauthoryear{De~Luca, Hang, Brudy, Lindner, and
  Hussmann}{De~Luca et~al\mbox{.}}{2012}]%
        {deluca2012touchme}
\bibfield{author}{\bibinfo{person}{Alexander De~Luca}, \bibinfo{person}{Alina
  Hang}, \bibinfo{person}{Frederik Brudy}, \bibinfo{person}{Christian Lindner},
  {and} \bibinfo{person}{Heinrich Hussmann}.} \bibinfo{year}{2012}\natexlab{}.
\newblock \showarticletitle{Touch Me Once and I Know It's You!: Implicit
  Authentication Based on Touch Screen Patterns}. In \bibinfo{booktitle}{{\em
  Proceedings of the SIGCHI Conference on Human Factors in Computing Systems}}
  {\em (\bibinfo{series}{CHI '12})}. \bibinfo{publisher}{ACM},
  \bibinfo{address}{New York, NY, USA}, \bibinfo{pages}{987--996}.
\newblock
\showISBNx{978-1-4503-1015-4}
\showDOI{%
\url{https://doi.org/10.1145/2207676.2208544}}


\bibitem[\protect\citeauthoryear{De~Luca, Harbach, von Zezschwitz, Maurer,
  Slawik, Hussmann, and Smith}{De~Luca et~al\mbox{.}}{2014}]%
        {deluca2014nowyouseeme}
\bibfield{author}{\bibinfo{person}{Alexander De~Luca}, \bibinfo{person}{Marian
  Harbach}, \bibinfo{person}{Emanuel von Zezschwitz},
  \bibinfo{person}{Max-Emanuel Maurer}, \bibinfo{person}{Bernhard~Ewald
  Slawik}, \bibinfo{person}{Heinrich Hussmann}, {and} \bibinfo{person}{Matthew
  Smith}.} \bibinfo{year}{2014}\natexlab{}.
\newblock \showarticletitle{Now You See Me, Now You Don'T: Protecting
  Smartphone Authentication from Shoulder Surfers}. In \bibinfo{booktitle}{{\em
  Proceedings of the 32Nd Annual ACM Conference on Human Factors in Computing
  Systems}} {\em (\bibinfo{series}{CHI '14})}. \bibinfo{publisher}{ACM},
  \bibinfo{address}{New York, NY, USA}, \bibinfo{pages}{2937--2946}.
\newblock
\showISBNx{978-1-4503-2473-1}
\showDOI{%
\url{https://doi.org/10.1145/2556288.2557097}}


\bibitem[\protect\citeauthoryear{De~Luca, Hertzschuch, and Hussmann}{De~Luca
  et~al\mbox{.}}{2010}]%
        {deluca2010colorpin}
\bibfield{author}{\bibinfo{person}{Alexander De~Luca}, \bibinfo{person}{Katja
  Hertzschuch}, {and} \bibinfo{person}{Heinrich Hussmann}.}
  \bibinfo{year}{2010}\natexlab{}.
\newblock \showarticletitle{ColorPIN: Securing PIN Entry Through Indirect
  Input}. In \bibinfo{booktitle}{{\em Proceedings of the SIGCHI Conference on
  Human Factors in Computing Systems}} {\em (\bibinfo{series}{CHI '10})}.
  \bibinfo{publisher}{ACM}, \bibinfo{address}{New York, NY, USA},
  \bibinfo{pages}{1103--1106}.
\newblock
\showISBNx{978-1-60558-929-9}
\showDOI{%
\url{https://doi.org/10.1145/1753326.1753490}}


\bibitem[\protect\citeauthoryear{De~Luca, Von~Zezschwitz, and
  Hu{\ss}mann}{De~Luca et~al\mbox{.}}{2009}]%
        {de2009vibrapass}
\bibfield{author}{\bibinfo{person}{Alexander De~Luca}, \bibinfo{person}{Emanuel
  Von~Zezschwitz}, {and} \bibinfo{person}{Heinrich Hu{\ss}mann}.}
  \bibinfo{year}{2009}\natexlab{}.
\newblock \showarticletitle{Vibrapass: secure authentication based on shared
  lies}. In \bibinfo{booktitle}{{\em Proceedings of the SIGCHI conference on
  human factors in computing systems}}. ACM, \bibinfo{pages}{913--916}.
\newblock


\bibitem[\protect\citeauthoryear{Eardley, Roudaut, Gill, and Thompson}{Eardley
  et~al\mbox{.}}{2017}]%
        {eardley2017grip}
\bibfield{author}{\bibinfo{person}{Rachel Eardley}, \bibinfo{person}{Anne
  Roudaut}, \bibinfo{person}{Steve Gill}, {and} \bibinfo{person}{Stephen~J.
  Thompson}.} \bibinfo{year}{2017}\natexlab{}.
\newblock \showarticletitle{Understanding Grip Shifts: How Form Factors Impact
  Hand Movements on Mobile Phones}. In \bibinfo{booktitle}{{\em Proceedings of
  the 2017 CHI Conference on Human Factors in Computing Systems}} {\em
  (\bibinfo{series}{CHI '17})}. \bibinfo{publisher}{ACM}, \bibinfo{address}{New
  York, NY, USA}, \bibinfo{pages}{4680--4691}.
\newblock
\showISBNx{978-1-4503-4655-9}
\showDOI{%
\url{https://doi.org/10.1145/3025453.3025835}}


\bibitem[\protect\citeauthoryear{Egelman, Jain, Portnoff, Liao, Consolvo, and
  Wagner}{Egelman et~al\mbox{.}}{2014}]%
        {egelman2014areyouready}
\bibfield{author}{\bibinfo{person}{Serge Egelman}, \bibinfo{person}{Sakshi
  Jain}, \bibinfo{person}{Rebecca~S. Portnoff}, \bibinfo{person}{Kerwell Liao},
  \bibinfo{person}{Sunny Consolvo}, {and} \bibinfo{person}{David Wagner}.}
  \bibinfo{year}{2014}\natexlab{}.
\newblock \showarticletitle{Are You Ready to Lock?}. In
  \bibinfo{booktitle}{{\em Proceedings of the 2014 ACM SIGSAC Conference on
  Computer and Communications Security}} {\em (\bibinfo{series}{CCS '14})}.
  \bibinfo{publisher}{ACM}, \bibinfo{address}{New York, NY, USA},
  \bibinfo{pages}{750--761}.
\newblock
\showISBNx{978-1-4503-2957-6}
\showDOI{%
\url{https://doi.org/10.1145/2660267.2660273}}


\bibitem[\protect\citeauthoryear{Farivar}{Farivar}{2015}]%
        {iphone6digit}
\bibfield{author}{\bibinfo{person}{Cyrus Farivar}.} \bibinfo{year}{Jun 8,
  2015}\natexlab{}.
\newblock \bibinfo{title}{Apple to require 6-digit passcodes on newer iPhones,
  iPads under iOS 9:Stronger passcode ups the ante: there will be one million
  possible permutations.}
\newblock   (\bibinfo{year}{Jun 8, 2015}).
\newblock
\newblock
\shownote{\url{http://arstechnica.com/apple/2015/06/apple-to-require-6-digit-passcodes-on-newer-iphones-ipads-under-ios-9/}.}


\bibitem[\protect\citeauthoryear{Forget, Chiasson, and Biddle}{Forget
  et~al\mbox{.}}{2010}]%
        {forget2010eyegaze}
\bibfield{author}{\bibinfo{person}{Alain Forget}, \bibinfo{person}{Sonia
  Chiasson}, {and} \bibinfo{person}{Robert Biddle}.}
  \bibinfo{year}{2010}\natexlab{}.
\newblock \showarticletitle{Shoulder-surfing Resistance with Eye-gaze Entry in
  Cued-recall Graphical Passwords}. In \bibinfo{booktitle}{{\em Proceedings of
  the SIGCHI Conference on Human Factors in Computing Systems}} {\em
  (\bibinfo{series}{CHI '10})}. \bibinfo{publisher}{ACM}, \bibinfo{address}{New
  York, NY, USA}, \bibinfo{pages}{1107--1110}.
\newblock
\showISBNx{978-1-60558-929-9}
\showDOI{%
\url{https://doi.org/10.1145/1753326.1753491}}


\bibitem[\protect\citeauthoryear{Gao, Ren, Chang, Liu, and Aickelin}{Gao
  et~al\mbox{.}}{2010}]%
        {gao2010new}
\bibfield{author}{\bibinfo{person}{H. Gao}, \bibinfo{person}{Z. Ren},
  \bibinfo{person}{X. Chang}, \bibinfo{person}{X. Liu}, {and}
  \bibinfo{person}{U. Aickelin}.} \bibinfo{year}{2010}\natexlab{}.
\newblock \showarticletitle{A New Graphical Password Scheme Resistant to
  Shoulder-Surfing}. In \bibinfo{booktitle}{{\em 2010 International Conference
  on Cyberworlds}}. \bibinfo{pages}{194--199}.
\newblock
\showDOI{%
\url{https://doi.org/10.1109/CW.2010.34}}


\bibitem[\protect\citeauthoryear{Harbach, De~Luca, and Egelman}{Harbach
  et~al\mbox{.}}{2016}]%
        {harbach2016anatomy}
\bibfield{author}{\bibinfo{person}{Marian Harbach}, \bibinfo{person}{Alexander
  De~Luca}, {and} \bibinfo{person}{Serge Egelman}.}
  \bibinfo{year}{2016}\natexlab{}.
\newblock \showarticletitle{The Anatomy of Smartphone Unlocking: A Field Study
  of Android Lock Screens}. In \bibinfo{booktitle}{{\em Proceedings of the 2016
  CHI Conference on Human Factors in Computing Systems}} {\em
  (\bibinfo{series}{CHI '16})}. \bibinfo{publisher}{ACM}, \bibinfo{address}{New
  York, NY, USA}, \bibinfo{pages}{4806--4817}.
\newblock
\showISBNx{978-1-4503-3362-7}
\showDOI{%
\url{https://doi.org/10.1145/2858036.2858267}}


\bibitem[\protect\citeauthoryear{Kelley, Komanduri, Mazurek, Shay, Vidas,
  Bauer, Christin, Cranor, and Lopez}{Kelley et~al\mbox{.}}{2012}]%
        {SP:KKMSVB12}
\bibfield{author}{\bibinfo{person}{Patrick~Gage Kelley},
  \bibinfo{person}{Saranga Komanduri}, \bibinfo{person}{Michelle~L. Mazurek},
  \bibinfo{person}{Richard Shay}, \bibinfo{person}{Timothy Vidas},
  \bibinfo{person}{Lujo Bauer}, \bibinfo{person}{Nicolas Christin},
  \bibinfo{person}{Lorrie~Faith Cranor}, {and} \bibinfo{person}{Julio Lopez}.}
  \bibinfo{year}{2012}\natexlab{}.
\newblock \showarticletitle{Guess Again (and Again and Again): Measuring
  Password Strength by Simulating Password-Cracking Algorithms}.
  \bibinfo{pages}{523--537}.
\newblock


\bibitem[\protect\citeauthoryear{Kim, Kim, Kim, and Cho}{Kim
  et~al\mbox{.}}{2011}]%
        {kim2011new}
\bibfield{author}{\bibinfo{person}{Sung-Hwan Kim}, \bibinfo{person}{Jong-Woo
  Kim}, \bibinfo{person}{Seon-Yeong Kim}, {and} \bibinfo{person}{Hwan-Gue
  Cho}.} \bibinfo{year}{2011}\natexlab{}.
\newblock \showarticletitle{A new shoulder-surfing resistant password for
  mobile environments}. In \bibinfo{booktitle}{{\em Proceedings of the 5th
  International Conference on Ubiquitous Information Management and
  Communication}}. ACM, \bibinfo{pages}{27}.
\newblock


\bibitem[\protect\citeauthoryear{Krombholz, Hupperich, and Holz}{Krombholz
  et~al\mbox{.}}{2017}]%
        {krombholz2017may}
\bibfield{author}{\bibinfo{person}{Katharina Krombholz},
  \bibinfo{person}{Thomas Hupperich}, {and} \bibinfo{person}{Thorsten Holz}.}
  \bibinfo{year}{2017}\natexlab{}.
\newblock \showarticletitle{May the Force Be with You: The Future of
  Force-Sensitive Authentication}.
\newblock \bibinfo{journal}{{\em IEEE Internet Computing\/}}
  \bibinfo{volume}{21}, \bibinfo{number}{3} (\bibinfo{year}{2017}),
  \bibinfo{pages}{64--69}.
\newblock


\bibitem[\protect\citeauthoryear{Kumar, Garfinkel, Boneh, and Winograd}{Kumar
  et~al\mbox{.}}{2007}]%
        {kumar2007reducing}
\bibfield{author}{\bibinfo{person}{Manu Kumar}, \bibinfo{person}{Tal
  Garfinkel}, \bibinfo{person}{Dan Boneh}, {and} \bibinfo{person}{Terry
  Winograd}.} \bibinfo{year}{2007}\natexlab{}.
\newblock \showarticletitle{Reducing Shoulder-surfing by Using Gaze-based
  Password Entry}. In \bibinfo{booktitle}{{\em Proceedings of the 3rd Symposium
  on Usable Privacy and Security}} {\em (\bibinfo{series}{SOUPS '07})}.
  \bibinfo{publisher}{ACM}, \bibinfo{address}{New York, NY, USA},
  \bibinfo{pages}{13--19}.
\newblock
\showISBNx{978-1-59593-801-5}
\showDOI{%
\url{https://doi.org/10.1145/1280680.1280683}}


\bibitem[\protect\citeauthoryear{Kwon, Shin, and Na}{Kwon
  et~al\mbox{.}}{2014}]%
        {kwon2014covert}
\bibfield{author}{\bibinfo{person}{Taekyoung Kwon}, \bibinfo{person}{Sooyeon
  Shin}, {and} \bibinfo{person}{Sarang Na}.} \bibinfo{year}{2014}\natexlab{}.
\newblock \showarticletitle{Covert attentional shoulder surfing: Human
  adversaries are more powerful than expected}.
\newblock \bibinfo{journal}{{\em IEEE Transactions on Systems, Man, and
  Cybernetics: Systems\/}} \bibinfo{volume}{44}, \bibinfo{number}{6}
  (\bibinfo{year}{2014}), \bibinfo{pages}{716--727}.
\newblock


\bibitem[\protect\citeauthoryear{Malek, Orozco, and El~Saddik}{Malek
  et~al\mbox{.}}{2006}]%
        {malek2006novel}
\bibfield{author}{\bibinfo{person}{Behzad Malek}, \bibinfo{person}{Mauricio
  Orozco}, {and} \bibinfo{person}{Abdulmotaleb El~Saddik}.}
  \bibinfo{year}{2006}\natexlab{}.
\newblock \showarticletitle{Novel shoulder-surfing resistant haptic-based
  graphical password}. In \bibinfo{booktitle}{{\em Proc. EuroHaptics}},
  Vol.~\bibinfo{volume}{6}.
\newblock


\bibitem[\protect\citeauthoryear{Man, Hong, and Matthews}{Man
  et~al\mbox{.}}{2003}]%
        {man2003shoulder}
\bibfield{author}{\bibinfo{person}{Shushuang Man}, \bibinfo{person}{Dawei
  Hong}, {and} \bibinfo{person}{Manton~M Matthews}.}
  \bibinfo{year}{2003}\natexlab{}.
\newblock \bibinfo{title}{A Shoulder-Surfing Resistant Graphical Password
  Scheme-WIW.}
\newblock   (\bibinfo{year}{2003}), \bibinfo{numpages}{105--111}~pages.
\newblock


\bibitem[\protect\citeauthoryear{Melicher, Kurilova, Segreti, Kalvani, Shay,
  Ur, Bauer, Christin, Cranor, and Mazurek}{Melicher et~al\mbox{.}}{2016}]%
        {melicher2016mobile}
\bibfield{author}{\bibinfo{person}{William Melicher}, \bibinfo{person}{Darya
  Kurilova}, \bibinfo{person}{Sean~M. Segreti}, \bibinfo{person}{Pranshu
  Kalvani}, \bibinfo{person}{Richard Shay}, \bibinfo{person}{Blase Ur},
  \bibinfo{person}{Lujo Bauer}, \bibinfo{person}{Nicolas Christin},
  \bibinfo{person}{Lorrie~Faith Cranor}, {and} \bibinfo{person}{Michelle~L.
  Mazurek}.} \bibinfo{year}{2016}\natexlab{}.
\newblock \showarticletitle{Usability and Security of Text Passwords on Mobile
  Devices}. In \bibinfo{booktitle}{{\em Proceedings of the 2016 CHI Conference
  on Human Factors in Computing Systems}} {\em (\bibinfo{series}{CHI '16})}.
\newblock


\bibitem[\protect\citeauthoryear{Roth, Richter, and Freidinger}{Roth
  et~al\mbox{.}}{2004}]%
        {roth2004pin}
\bibfield{author}{\bibinfo{person}{Volker Roth}, \bibinfo{person}{Kai Richter},
  {and} \bibinfo{person}{Rene Freidinger}.} \bibinfo{year}{2004}\natexlab{}.
\newblock \showarticletitle{A PIN-entry method resilient against shoulder
  surfing}. In \bibinfo{booktitle}{{\em Proceedings of the 11th ACM conference
  on Computer and communications security}}. ACM, \bibinfo{pages}{236--245}.
\newblock


\bibitem[\protect\citeauthoryear{Sahami~Shirazi, Moghadam, Ketabdar, and
  Schmidt}{Sahami~Shirazi et~al\mbox{.}}{2012}]%
        {sahami2012assessing}
\bibfield{author}{\bibinfo{person}{Alireza Sahami~Shirazi},
  \bibinfo{person}{Peyman Moghadam}, \bibinfo{person}{Hamed Ketabdar}, {and}
  \bibinfo{person}{Albrecht Schmidt}.} \bibinfo{year}{2012}\natexlab{}.
\newblock \showarticletitle{Assessing the vulnerability of magnetic gestural
  authentication to video-based shoulder surfing attacks}. In
  \bibinfo{booktitle}{{\em Proceedings of the SIGCHI Conference on Human
  Factors in Computing Systems}}. ACM, \bibinfo{pages}{2045--2048}.
\newblock


\bibitem[\protect\citeauthoryear{Schaub, Deyhle, and Weber}{Schaub
  et~al\mbox{.}}{2012}]%
        {schaub2012password}
\bibfield{author}{\bibinfo{person}{Florian Schaub}, \bibinfo{person}{Ruben
  Deyhle}, {and} \bibinfo{person}{Michael Weber}.}
  \bibinfo{year}{2012}\natexlab{}.
\newblock \showarticletitle{Password Entry Usability and Shoulder Surfing
  Susceptibility on Different Smartphone Platforms}. In
  \bibinfo{booktitle}{{\em Proceedings of the 11th International Conference on
  Mobile and Ubiquitous Multimedia}} {\em (\bibinfo{series}{MUM '12})}.
  \bibinfo{publisher}{ACM}, \bibinfo{address}{New York, NY, USA}, Article
  \bibinfo{articleno}{13}, \bibinfo{numpages}{10}~pages.
\newblock
\showISBNx{978-1-4503-1815-0}
\showDOI{%
\url{https://doi.org/10.1145/2406367.2406384}}


\bibitem[\protect\citeauthoryear{Schaub, Walch, K{\"o}nings, and Weber}{Schaub
  et~al\mbox{.}}{2013}]%
        {schaub2013exploring}
\bibfield{author}{\bibinfo{person}{Florian Schaub}, \bibinfo{person}{Marcel
  Walch}, \bibinfo{person}{Bastian K{\"o}nings}, {and} \bibinfo{person}{Michael
  Weber}.} \bibinfo{year}{2013}\natexlab{}.
\newblock \showarticletitle{Exploring the design space of graphical passwords
  on smartphones}. In \bibinfo{booktitle}{{\em Proceedings of the Ninth
  Symposium on Usable Privacy and Security}}. ACM, \bibinfo{pages}{11}.
\newblock


\bibitem[\protect\citeauthoryear{Sherman, Clark, Yang, Sugrim, Modig,
  Lindqvist, Oulasvirta, and Roos}{Sherman et~al\mbox{.}}{2014}]%
        {sherman2014user}
\bibfield{author}{\bibinfo{person}{Michael Sherman}, \bibinfo{person}{Gradeigh
  Clark}, \bibinfo{person}{Yulong Yang}, \bibinfo{person}{Shridatt Sugrim},
  \bibinfo{person}{Arttu Modig}, \bibinfo{person}{Janne Lindqvist},
  \bibinfo{person}{Antti Oulasvirta}, {and} \bibinfo{person}{Teemu Roos}.}
  \bibinfo{year}{2014}\natexlab{}.
\newblock \showarticletitle{User-generated free-form gestures for
  authentication: Security and memorability}. In \bibinfo{booktitle}{{\em
  Proceedings of the 12th annual international conference on Mobile systems,
  applications, and services}}. ACM, \bibinfo{pages}{176--189}.
\newblock


\bibitem[\protect\citeauthoryear{Tipton, White~II, Sershon, and Choi}{Tipton
  et~al\mbox{.}}{2014}]%
        {tipton2014ios}
\bibfield{author}{\bibinfo{person}{Stephen~J Tipton}, \bibinfo{person}{Daniel~J
  White~II}, \bibinfo{person}{Christopher Sershon}, {and}
  \bibinfo{person}{Young~B Choi}.} \bibinfo{year}{2014}\natexlab{}.
\newblock \showarticletitle{iOS security and privacy: Authentication methods,
  permissions, and potential pitfalls with touch id}.
\newblock \bibinfo{journal}{{\em International Journal of Computer and
  Information Technology\/}} \bibinfo{volume}{3}, \bibinfo{number}{03}
  (\bibinfo{year}{2014}).
\newblock


\bibitem[\protect\citeauthoryear{Uellenbeck, D\"{u}rmuth, Wolf, and
  Holz}{Uellenbeck et~al\mbox{.}}{2013}]%
        {uellenbeck}
\bibfield{author}{\bibinfo{person}{Sebastian Uellenbeck},
  \bibinfo{person}{Markus D\"{u}rmuth}, \bibinfo{person}{Christopher Wolf},
  {and} \bibinfo{person}{Thorsten Holz}.} \bibinfo{year}{2013}\natexlab{}.
\newblock \showarticletitle{Quantifying the Security of Graphical Passwords:
  The Case of Android Unlock Patterns}. In \bibinfo{booktitle}{{\em Proceedings
  of the 2013 ACM SIGSAC Conference on Computer \&\#38; Communications
  Security}} {\em (\bibinfo{series}{CCS '13})}. \bibinfo{publisher}{ACM},
  \bibinfo{address}{New York, NY, USA}, \bibinfo{pages}{161--172}.
\newblock
\showISBNx{978-1-4503-2477-9}
\showDOI{%
\url{https://doi.org/10.1145/2508859.2516700}}


\bibitem[\protect\citeauthoryear{Van~Nguyen, Sae-Bae, and Memon}{Van~Nguyen
  et~al\mbox{.}}{2017}]%
        {van2017draw}
\bibfield{author}{\bibinfo{person}{Toan Van~Nguyen}, \bibinfo{person}{Napa
  Sae-Bae}, {and} \bibinfo{person}{Nasir Memon}.}
  \bibinfo{year}{2017}\natexlab{}.
\newblock \showarticletitle{DRAW-A-PIN: Authentication using finger-drawn PIN
  on touch devices}.
\newblock \bibinfo{journal}{{\em Computers \& Security\/}}
  \bibinfo{volume}{66} (\bibinfo{year}{2017}), \bibinfo{pages}{115--128}.
\newblock


\bibitem[\protect\citeauthoryear{Von~Zezschwitz, De~Luca, Brunkow, and
  Hussmann}{Von~Zezschwitz et~al\mbox{.}}{2015}]%
        {von2015swipin}
\bibfield{author}{\bibinfo{person}{Emanuel Von~Zezschwitz},
  \bibinfo{person}{Alexander De~Luca}, \bibinfo{person}{Bruno Brunkow}, {and}
  \bibinfo{person}{Heinrich Hussmann}.} \bibinfo{year}{2015}\natexlab{}.
\newblock \showarticletitle{SwiPIN: Fast and secure pin-entry on smartphones}.
  In \bibinfo{booktitle}{{\em Proceedings of the 33rd Annual ACM Conference on
  Human Factors in Computing Systems}}. ACM, \bibinfo{pages}{1403--1406}.
\newblock


\bibitem[\protect\citeauthoryear{von Zezschwitz, De~Luca, Janssen, and
  Hussmann}{von Zezschwitz et~al\mbox{.}}{2015}]%
        {vzw2015easy}
\bibfield{author}{\bibinfo{person}{Emanuel von Zezschwitz},
  \bibinfo{person}{Alexander De~Luca}, \bibinfo{person}{Philipp Janssen}, {and}
  \bibinfo{person}{Heinrich Hussmann}.} \bibinfo{year}{2015}\natexlab{}.
\newblock \showarticletitle{Easy to Draw, but Hard to Trace?: On the
  Observability of Grid-based (Un)Lock Patterns}. In \bibinfo{booktitle}{{\em
  Proceedings of the 33rd Annual ACM Conference on Human Factors in Computing
  Systems}} {\em (\bibinfo{series}{CHI '15})}. \bibinfo{publisher}{ACM},
  \bibinfo{address}{New York, NY, USA}, \bibinfo{pages}{2339--2342}.
\newblock
\showISBNx{978-1-4503-3145-6}
\showDOI{%
\url{https://doi.org/10.1145/2702123.2702202}}


\bibitem[\protect\citeauthoryear{von Zezschwitz, Dunphy, and De~Luca}{von
  Zezschwitz et~al\mbox{.}}{2013}]%
        {vonZezschwitz2013wild}
\bibfield{author}{\bibinfo{person}{Emanuel von Zezschwitz},
  \bibinfo{person}{Paul Dunphy}, {and} \bibinfo{person}{Alexander De~Luca}.}
  \bibinfo{year}{2013}\natexlab{}.
\newblock \showarticletitle{Patterns in the Wild: A Field Study of the
  Usability of Pattern and {PIN}-based Authentication on Mobile Devices}. In
  \bibinfo{booktitle}{{\em Proceedings of the 15th International Conference on
  Human-computer Interaction with Mobile Devices and Services}} {\em
  (\bibinfo{series}{MobileHCI '13})}. \bibinfo{pages}{261--270}.
\newblock


\bibitem[\protect\citeauthoryear{von Zezschwitz, Eiband, Buschek, Oberhuber,
  De~Luca, Alt, and Hussmann}{von Zezschwitz et~al\mbox{.}}{2016}]%
        {vonZezschwitz2016quant}
\bibfield{author}{\bibinfo{person}{Emanuel von Zezschwitz},
  \bibinfo{person}{Malin Eiband}, \bibinfo{person}{Daniel Buschek},
  \bibinfo{person}{Sascha Oberhuber}, \bibinfo{person}{Alexander De~Luca},
  \bibinfo{person}{Florian Alt}, {and} \bibinfo{person}{Heinrich Hussmann}.}
  \bibinfo{year}{2016}\natexlab{}.
\newblock \showarticletitle{On Quantifying the Effective Password Space of
  Grid-based Unlock Gestures}. In \bibinfo{booktitle}{{\em Proceedings of the
  15th International Conference on Mobile and Ubiquitous Multimedia}} {\em
  (\bibinfo{series}{MUM '16})}. \bibinfo{publisher}{ACM}, \bibinfo{address}{New
  York, NY, USA}, \bibinfo{pages}{201--212}.
\newblock
\showISBNx{978-1-4503-4860-7}
\showDOI{%
\url{https://doi.org/10.1145/3012709.3012729}}


\bibitem[\protect\citeauthoryear{Weir, Aggarwal, Collins, and Stern}{Weir
  et~al\mbox{.}}{2010}]%
        {CCS:WACS10}
\bibfield{author}{\bibinfo{person}{Matt Weir}, \bibinfo{person}{Sudhir
  Aggarwal}, \bibinfo{person}{Michael Collins}, {and} \bibinfo{person}{Henry
  Stern}.} \bibinfo{year}{2010}\natexlab{}.
\newblock \showarticletitle{Testing metrics for password creation policies by
  attacking large sets of revealed passwords}. \bibinfo{pages}{162--175}.
\newblock


\bibitem[\protect\citeauthoryear{Wiedenbeck, Waters, Sobrado, and
  Birget}{Wiedenbeck et~al\mbox{.}}{2006}]%
        {wiedenbeck2006design}
\bibfield{author}{\bibinfo{person}{Susan Wiedenbeck}, \bibinfo{person}{Jim
  Waters}, \bibinfo{person}{Leonardo Sobrado}, {and}
  \bibinfo{person}{Jean-Camille Birget}.} \bibinfo{year}{2006}\natexlab{}.
\newblock \showarticletitle{Design and Evaluation of a Shoulder-surfing
  Resistant Graphical Password Scheme}. In \bibinfo{booktitle}{{\em Proceedings
  of the Working Conference on Advanced Visual Interfaces}} {\em
  (\bibinfo{series}{AVI '06})}. \bibinfo{publisher}{ACM}, \bibinfo{address}{New
  York, NY, USA}, \bibinfo{pages}{177--184}.
\newblock
\showISBNx{1-59593-353-0}
\showDOI{%
\url{https://doi.org/10.1145/1133265.1133303}}


\bibitem[\protect\citeauthoryear{Wiese and Roth}{Wiese and Roth}{2015}]%
        {wiese2015pitfalls}
\bibfield{author}{\bibinfo{person}{Oliver Wiese} {and} \bibinfo{person}{Volker
  Roth}.} \bibinfo{year}{2015}\natexlab{}.
\newblock \showarticletitle{Pitfalls of Shoulder Surfing Studies}. In
  \bibinfo{booktitle}{{\em NDSS Workshop on Usable Security}}.
  \bibinfo{pages}{1--6}.
\newblock


\bibitem[\protect\citeauthoryear{Zakaria, Griffiths, Brostoff, and Yan}{Zakaria
  et~al\mbox{.}}{2011}]%
        {zakaria2011shoulder}
\bibfield{author}{\bibinfo{person}{Nur~Haryani Zakaria}, \bibinfo{person}{David
  Griffiths}, \bibinfo{person}{Sacha Brostoff}, {and} \bibinfo{person}{Jeff
  Yan}.} \bibinfo{year}{2011}\natexlab{}.
\newblock \showarticletitle{Shoulder surfing defence for recall-based graphical
  passwords}. In \bibinfo{booktitle}{{\em Proceedings of the Seventh Symposium
  on Usable Privacy and Security}}. ACM, \bibinfo{pages}{6}.
\newblock


\bibitem[\protect\citeauthoryear{Zhao, Ahn, Seo, and Hu}{Zhao
  et~al\mbox{.}}{2013}]%
        {zhao2013security}
\bibfield{author}{\bibinfo{person}{Ziming Zhao}, \bibinfo{person}{Gail-Joon
  Ahn}, \bibinfo{person}{Jeong-Jin Seo}, {and} \bibinfo{person}{Hongxin Hu}.}
  \bibinfo{year}{2013}\natexlab{}.
\newblock \showarticletitle{On the security of picture gesture authentication}.
  In \bibinfo{booktitle}{{\em 22nd USENIX Security Symposium (USENIX Security
  13)}}. \bibinfo{pages}{383--398}.
\newblock


\end{thebibliography}

\appendix
\counterwithin{figure}{section}

\section{Survey Advertisement}
\label{app:advert}

\begin{small}
  We are conducting an academic survey about shoulder surfing on
  mobile device authentication mechanisms. We would like you to act as
  an attacker attempting to get someone's mobile device password by
  observing videos of a user authenticating into a mobile device. If
  you are currently viewing this page on a mobile device (ie. cell
  phone or tablet), please switch to a desktop or laptop computer to
  take this survey. If you get to the survey and it detects a mobile
  device, you will be opted out of the survey. Please select the link
  below to complete the survey. At the end of the survey, you will
  receive a code to enter into the submission form below to receive
  credit for taking our survey. \\\\

THE SURVEY WILL ONLY WORK IF YOU VIEW IT ON A NON-MOBILE DEVICE COMPUTER.\\\\

We have only tested the survey using GOOGLE CHROME OR MOZILLA FIREFOX. If you experience problems, opt out and return the HIT without penalty.\\\\

You will be compensated \$1.50 for your work. We have found that it takes approximately 10 minutes on average to complete this HIT, for a payout of about \$0.15 a minute \\\\

Due to the nature of the work, you may only complete the HIT once, even across multiple posting of the HIT. If you accept the HIT and are notified that your work will not be accepted, please return the HIT. FAILURE TO FOLLOW THIS INSTRUCTION MAY RESULT IN WORK BEING EXCLUDED AND/OR A REJECTION.\\\\

Please feel free to contact the requester if you have any questions or concerns. A prompt reply should occur within 24 hours or sooner.\\\\

Note: this survey requires your browser to load several high quality videos. We do not recommend you attempt this survey if you have a limited data connection.\\\\
\end{small}
\section{Patterns and PINs Visualized}

\subsection{Patterns}
\label{fig:patterns}
Patterns used with properties: a double circle indicates a start point, while a single circle indicates a point included in the pattern. Note that labeling
of patterns begins in the upper left with 0, ending in the lower right
with 8 (See Figure~\ref{fig:samples})

\bigskip

\begin{center}
\begin{tabular}{c c}
\fbox{\includegraphics[width=0.25\linewidth]{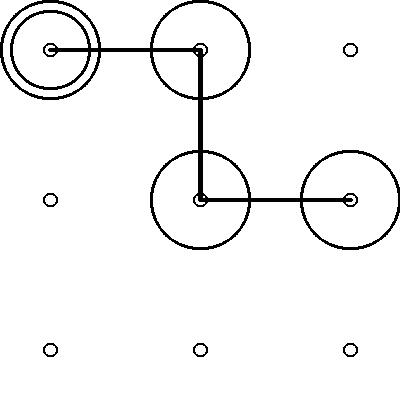}} &
\fbox{\includegraphics[width=0.25\linewidth]{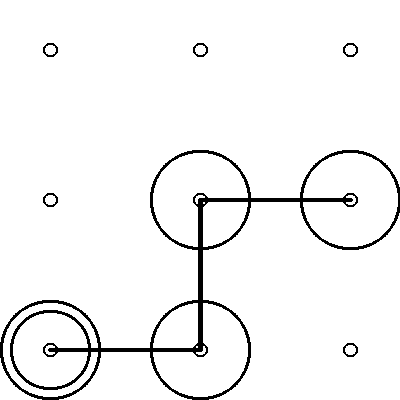}} \\
0145 & 6745 \\
up & down\\\\
\end{tabular}

\begin{tabular}{c c}
\fbox{\includegraphics[width=0.25\linewidth]{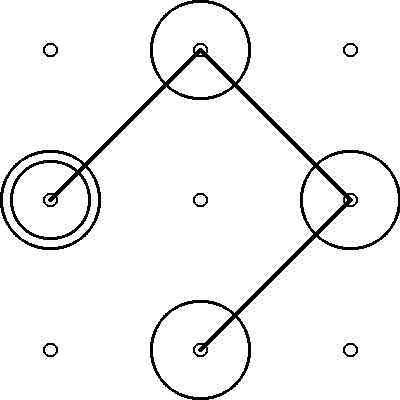}} &
\fbox{\includegraphics[width=0.25\linewidth]{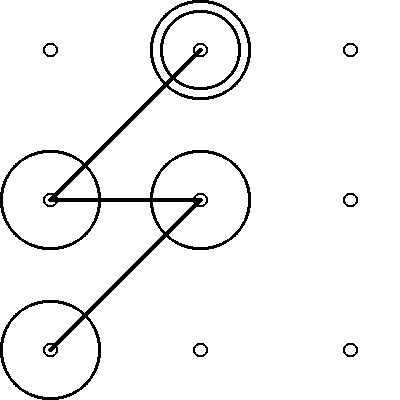}} \\
3157 &  1346 \\
 neutral & left \\\\
\end{tabular}

\begin{tabular}{c c}
\fbox{\includegraphics[width=0.25\linewidth]{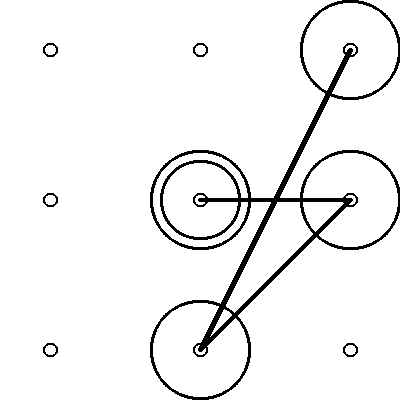}} & \fbox{\includegraphics[width=0.25\linewidth]{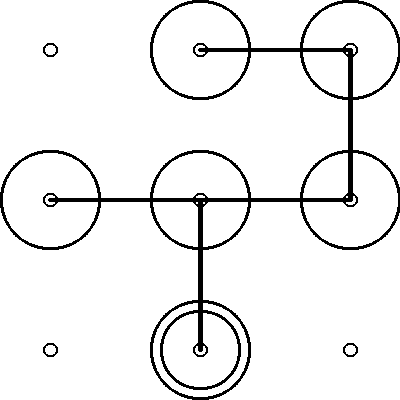}} \\
4572  & 743521  \\
right/cross & up/non-adj \\ \\
\end{tabular}

\begin{tabular}{c c}
\fbox{\includegraphics[width=0.25\linewidth]{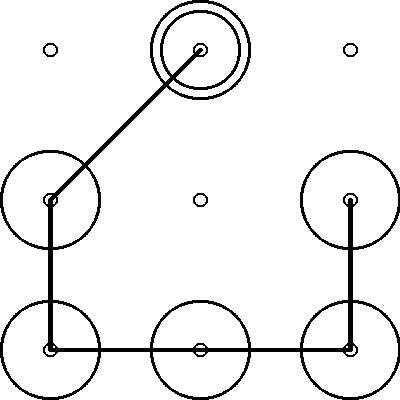}} &
\fbox{\includegraphics[width=0.25\linewidth]{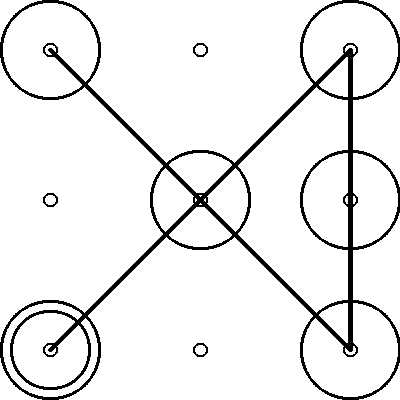}} \\
 136785 &  642580\\
 down & neutral/cross \\ \\
\end{tabular}

\begin{tabular}{c c}
\fbox{\includegraphics[width=0.25\linewidth]{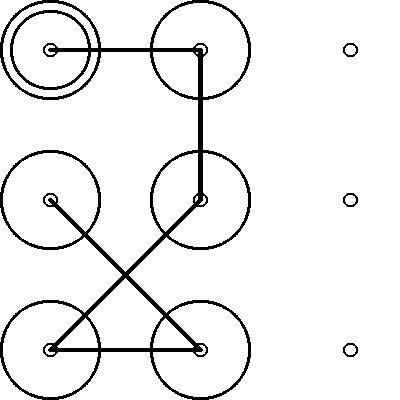}} &
\fbox{\includegraphics[width=0.25\linewidth]{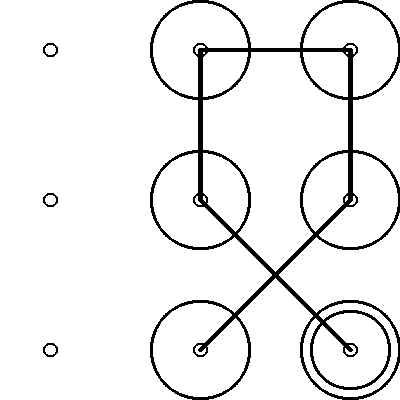}} \\
 014673  & 841257 \\
left & right/kmove/cross \\
\end{tabular}
\end{center}

\subsection{PINs}
\label{fig:pins}

PINs used with properties: filled circle is the start point, multiple circles on a number indicate multiple touches. 

\bigskip

\begin{center}
  \begin{tabular}{c c}
    \fbox{\includegraphics[width=0.25\linewidth]{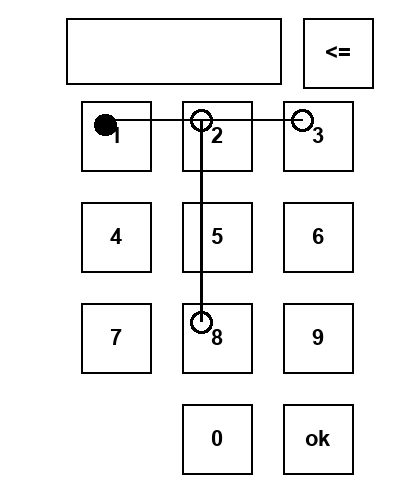}}&
    \fbox{\includegraphics[width=0.25\linewidth]{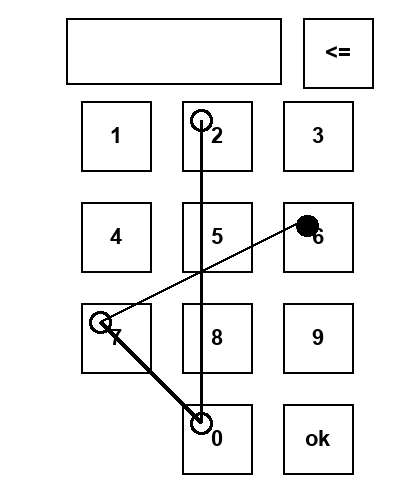}}\\
1328 & 6702 \\
up/non-adj   & down/kmove/cross\\\\
  \end{tabular}

  \begin{tabular}{c c}
    \fbox{\includegraphics[width=0.25\linewidth]{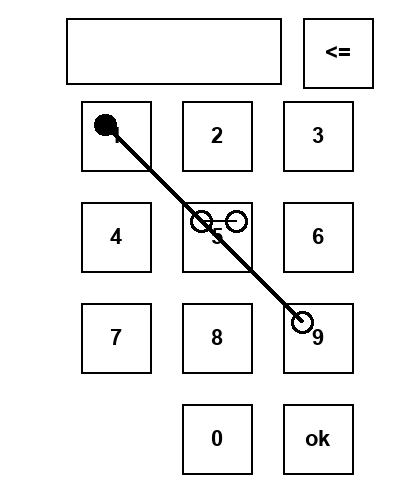}}&
    \fbox{\includegraphics[width=0.25\linewidth]{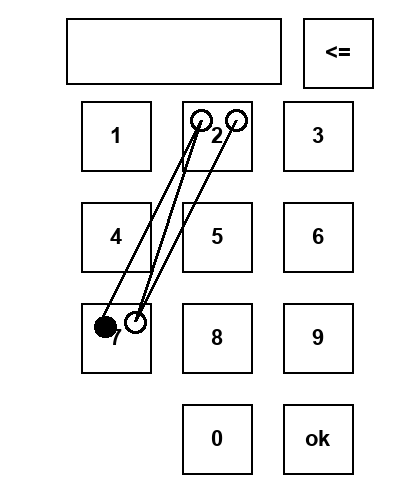}}\\
 1955    & 7272\\
 neutral/non-adj/repeats & left/kmoves/repeats\\\\
  \end{tabular}

  \begin{tabular}{c c}
    \fbox{\includegraphics[width=0.25\linewidth]{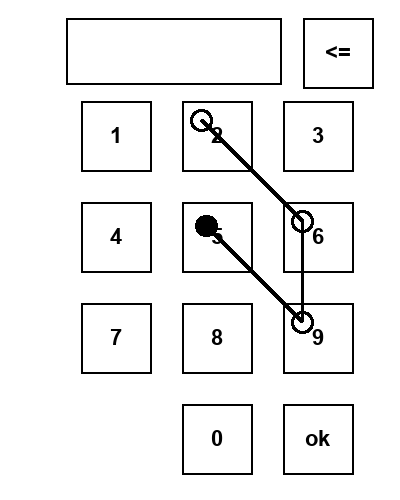}}&
    \fbox{\includegraphics[width=0.25\linewidth]{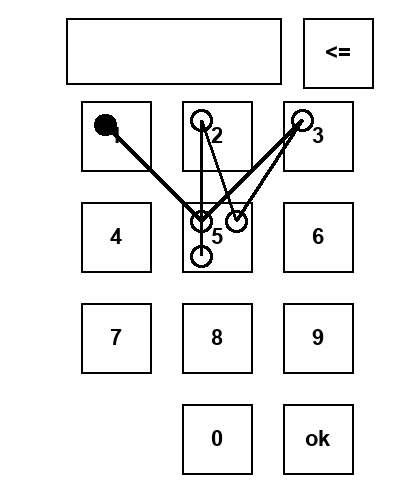}}\\
 5962 & 152525\\
 right & up/repat \\\\
  \end{tabular}

  \begin{tabular}{c c}
    \fbox{\includegraphics[width=0.25\linewidth]{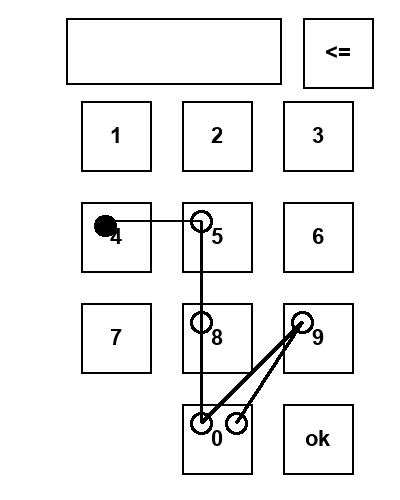}}&
    \fbox{\includegraphics[width=0.25\linewidth]{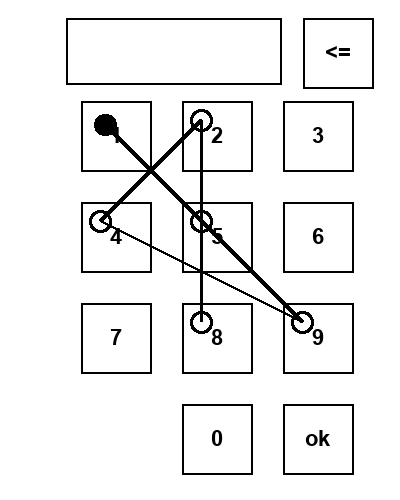}}\\
 458090 & 159428 \\
 down/repeat & neutral/cross/non-adj\\\\
  \end{tabular}

  \begin{tabular}{c c}
    \fbox{\includegraphics[width=0.25\linewidth]{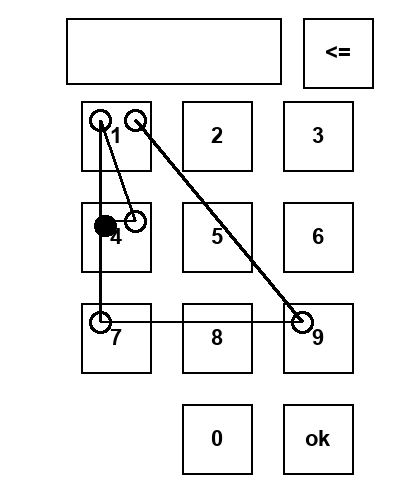}}&
    \fbox{\includegraphics[width=0.25\linewidth]{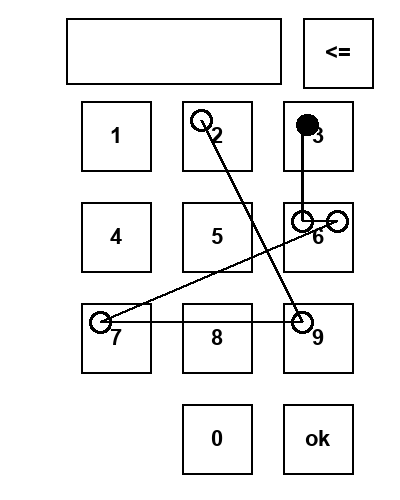}}\\
 441791 & 366792 \\
 left/kmove/repeat & right/repeat/kmove/cross\\
  \end{tabular}
\end{center}


\end{document}